\documentclass[12pt]{article}
\usepackage[a4paper, margin=2cm]{geometry}
\usepackage{authblk}

\usepackage{graphicx}
\usepackage{hyperref}
\usepackage{cite}

\usepackage{amsmath}
\usepackage{amssymb}

\usepackage[toc,page]{appendix}

\usepackage{parskip}
\usepackage{setspace}
\title{Investigation of Wake Dynamics of a Slender Symmetric Trailing Edge Hydrofoil}

\author{Gabriele Gaiti}
\author{Chirag Trivedi}
\author{Kristian F. Sagmo}

\affil{\small Waterpower Laboratory, Department of Energy and Process Engineering, NTNU — Norwegian University of Science and Technology, Alfred Getz' vei 4, 7034 Trondheim, Norway.}

\date{\small \today}

\begin{document}

\maketitle

\begin{abstract}
\noindent Accurate prediction of wake dynamics behind hydrofoils is critical for mitigating vortex-induced vibrations and improving the performance of hydraulic machinery. Conventional turbulence modeling approaches often struggle to capture the unsteady, coherent structures governing wake behavior, particularly for slender hydrofoils operating at high Reynolds numbers. This study addresses this limitation by combining scale-resolving numerical simulations, including high-resolution Large Eddy Simulation (LES), with Particle Image Velocimetry (PIV) measurements to investigate the turbulent wake of a symmetric, blunt trailing-edge hydrofoil operating at zero angle of attack. The flow was analyzed at a Reynolds number of approximately 7.5 $\times 10^5$, i.e. close to the onset of wake–structure interaction effects. LES was performed using a fine mesh of approximately 500 million nodes to resolve near-wall and wake dynamics beyond the experimental field of view, while PIV measurements provided time-resolved velocity fields downstream of the trailing edge. Proper Orthogonal Decomposition (POD) was applied to the PIV data to extract dominant coherent structures and quantify their contribution to the turbulent kinetic energy. POD analysis reveals that energy is distributed across many modes, with the leading mode capturing the primary wake dynamics and higher modes forming coupled oscillatory pairs associated with von Kármán vortex shedding. PIV–LES agreement shows that central wake measurements combined with numerical simulations enables full wake reconstruction and validates modeling for vibration-relevant hydrofoil dynamics.
\end{abstract}

\section{Introduction}
\begin{figure}[ht]
\centering
\includegraphics[width=0.6\columnwidth]{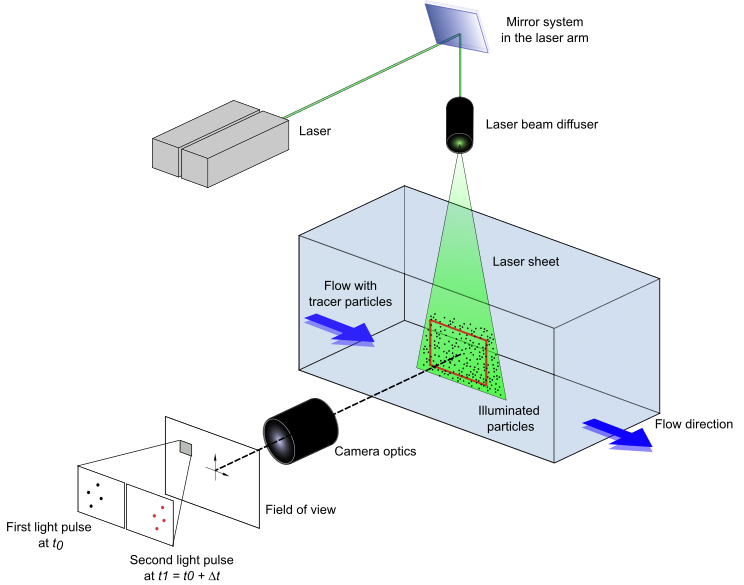}
\caption{Schematic view of the velocity measurement using PIV technique.}
\label{fig:PIVconcept}
\end{figure}

Understanding the behavior of physical elements at detailed scales has been the focus of researchers for quite some time. The ability to validate results obtained through simulations using experiments has proven to be critical for the validation of computational models, especially in the face of important approximations dependent on the turbulence model used \cite{temmerman_investigation_2003}. One of the most widely used methods for extracting fluid dynamic parameters from an experimental model is the Particle Image Velocimetry (PIV). This is an optical, indirect, and almost non-intrusive method that allows the calculation of the instantaneous velocity field with an accuracy that could reach sub-pixel level. Its functioning principles can be seen in Figure ~\ref{fig:PIVconcept}, and basically consists in tracking the displacement of a particle (or group of particles) over a specific time interval. Beyond conventional two-dimensional PIV, advanced tomographic implementations can provide high-resolution three-dimensional wake imaging, revealing complex vortical structures \cite{ziazi_tomographic_2019}. The fluid-structure interaction between submerged bodies and turbulent flow results in a composition of rather complex physical phenomena \cite{williamson_vortex_1996}. The present study focuses on turbulence, vortex shedding and coherent structures. As already discovered in 1960, the vortex shedding frequency of symmetrical-blunt hydrofoil is mainly dependent on turbulence, viscosity and boundary layer conditions \cite{heskestad_influence_1960}.

Literature have shown that hydrofoil trailing-edge geometry, planform variations, and passive flow-control devices such as vortex generators can significantly influence wake dynamics, vortex shedding, and vortex-induced vibrations, as well as alter tip-vortex roll-up and overall wake morphology, indicating that wake structure is sensitive to both trailing-edge features and global hydrofoil geometry \cite{zobeiri_effect_2012, chen_numerical_2020, besirovic_vortex_2020, sagmo_piv_2019, fruman_effect_1995}. Also cavitation and fluid–structure interaction significantly influence von Kármán vortex shedding characteristics and coherence in trailing-edge wakes, particularly under resonance conditions that amplify wake instability and fluctuating forces on the hydrofoil \cite{ausoni_cavitation_2007}. Turbulence modeling, particularly when using traditional Reynolds-averaged Navier–Stokes (RANS) approaches, often fails to capture the detailed, unsteady wake structures that drive such phenomena, thereby limiting predictive accuracy for wake behavior and VIV \cite{zobeiri_effect_2012, hu_numerical_2020, dauengauer_flow_2018}. Experimental investigations using high-resolution methods such as PIV have thus proven critical to reveal the true flow physics, including vortex shedding behavior, wake mixing, and wake-induced forces \cite{ben-gida_stratified_2016, zobeiri_effect_2012, chen_numerical_2020}. 
Researchers have investigated time-resolved PIV with LES decomposition for a flexible airfoil wake \cite{siala_characterization_2016}, quantifying energy deficits and coherent structures, highlighting the importance of resolving unsteady shedding both experimentally and numerically.

However, no publicly available studies combine both high-resolution PIV data and scale-resolving numerical simulations as LES for a slender hydrofoil geometry, in fixed-beam configuration, in water. To address this gap, the present study combines scale-resolving numerical simulations as LES with PIV measurements of comparable spatial resolution (0.1–0.35 mm in the simulations and approximately 0.55 mm in the experiments) providing a basis for assessing the ability of LES to reproduce the key turbulent structures present in the wake. In addition, Proper Orthogonal Decomposition (POD) is applied to the PIV velocity fields to extract the dominant coherent structures in an energy-ranked hierarchy. This modal decomposition is particularly effective in hydrofoil wakes, where periodic vortex shedding, shear-layer instabilities, and small-scale turbulent motions coexist. By isolating the large-scale energetic modes from background turbulence, POD offers a clearer physical interpretation of the unsteady wake dynamics. The combined PIV-POD–LES framework thereby provides a new, high-resolution experimental dataset for evaluating scale-resolving turbulence models, while deepening the understanding of wake dynamics relevant for vibration mitigation and slender symmetric hydrofoil performance optimization.

\section{Experimental Setup}
\label{sec:expsetup}
\begin{figure}[ht]
\centering
\includegraphics[width=0.5\columnwidth]{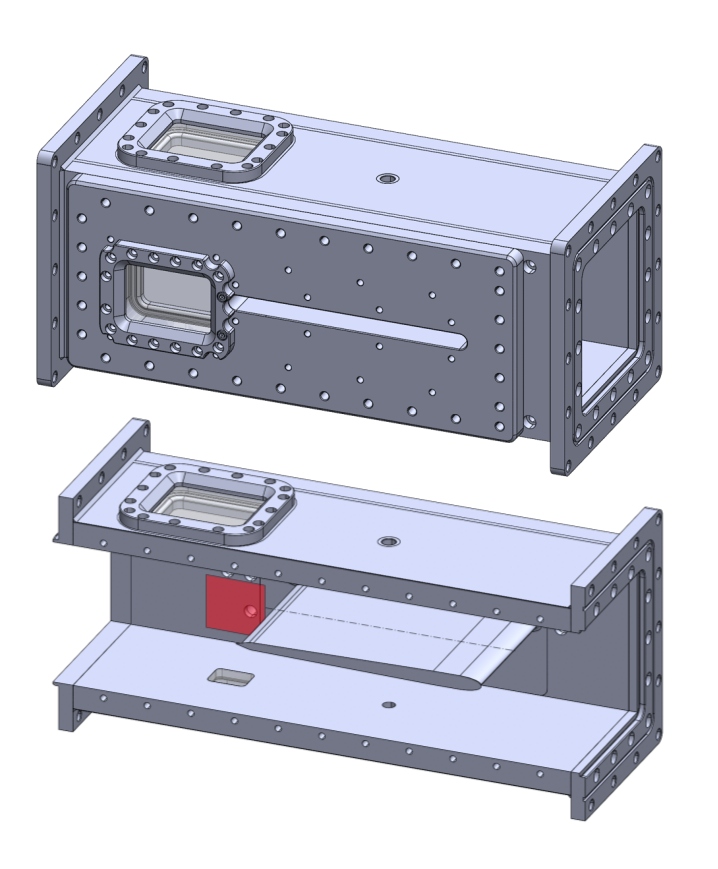}
\caption{View of test section (above) and section view (below) with PIV FOV in red. The frontal and top opening gave optical access to camera and laser sheet respectively.}
\label{fig:testsection}
\end{figure}

\begin{figure}[ht]
\centering
\includegraphics[width=0.5\columnwidth]{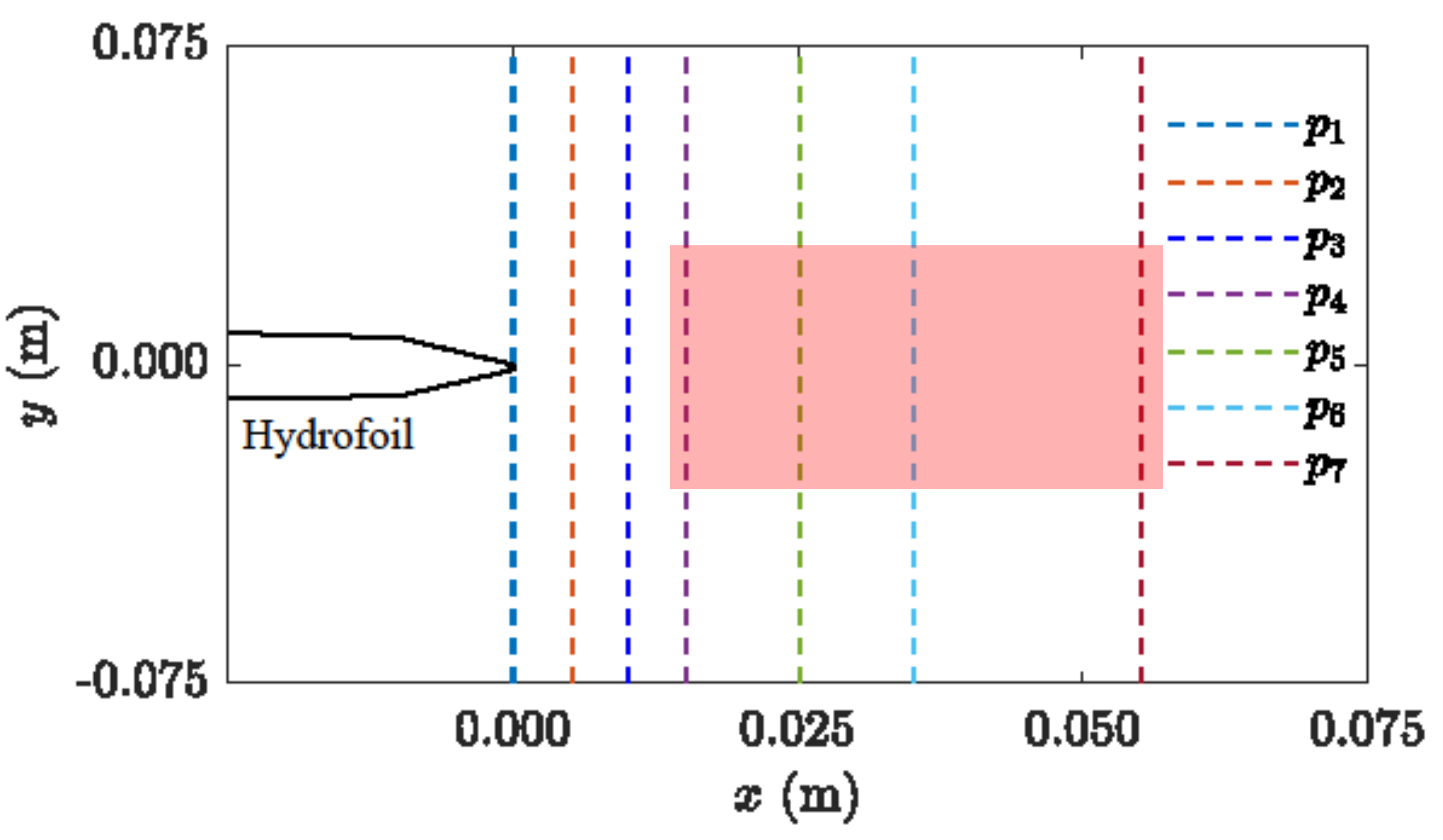}
\caption{Velocity profile locations downstream of the hydrofoil. The profile lines $p_1$, $p_2$, $p_3$, $p_4$, $p_5$, $p_6$, $p_7$, correspond to the chord length of 0\%, 2\%, 4\%, 6\%, 10\%, 14\% and 22\%, respectively. In red the area investigated by the PIV. \label{fig:velprofileloc}}
\end{figure}

The experiment has been performed at the Waterpower Laboratory of the Norwegian University of Science and Technology. The test-section setup and field of view (FOV) are illustrated in Figure \ref{fig:testsection}, the locations of the profile sections considered at various distances downstream of the foil trailing edge are shown in Figure \ref{fig:velprofileloc}. The detailed hydrofoil geometry is illustrated in Figure \ref{fig:foilmeas1}.

\subsection{Test section}
\begin{figure}[ht]
\centering
\includegraphics[width=0.5\columnwidth]{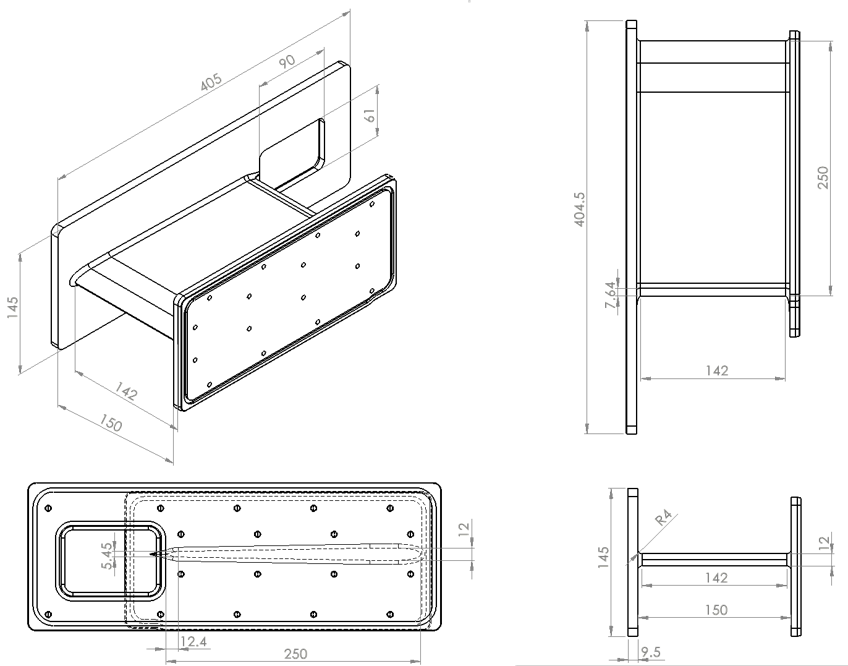}
\caption{Isometric view of the hydrofoil. All dimensions are in mm. }
\label{fig:foilmeas1}
\end{figure}

The experiments were conducted in close loop configuration, square channel (150x150 mm) and cavitation free environment by allowing a geostatic pressure of 6 mWc over the test section level. The hydrofoil resembles a high head Francis runner blade with symmetric blunt trailing edge and has been tested without angle of attack and in fixed-beam configuration. As can be seen from Figure \ref{fig:foilmeas1} the hydrofoil has been machined together with its supporting walls. The optical access has been given by transparent, flat plexiglass windows with a thickness of 40 mm. The FOV starts at a distance of 13.15 mm from the trailing edge, along the hydrofoil centerline. Within the FOV three sections of interest were analyzed in detail, as in Table \ref{tab:FOVdetails}, both in the PIV experiment and the LES simulation, as this region is where vortex detachment is expected for the specific velocity considered. In order to ensure statistical significance, intended as improved correlation of the results and reduction of outliers, a set of six PIV measurements has been carried out. Other important measurement parameters are described in the summary Table \ref{tab:PIVflowdetails}.

\begin{table}[ht]
\caption{FOV parameters\label{tab:FOVdetails}}
\centering{%
\begin{tabular}{l r}
\hline
FOV Dimensions&56.85 x 45 mm  \\
FOV distance from trailing edge &13.15 mm\\\\
Point of interest position from trailing edge:\\
   $p_5$ & 25.14 mm\\
   $p_6$ & 35.14 mm\\
   $p_7$ & 55.14 mm\\\\
   Position uncertainty & $\pm$ 0.5 mm \\
\hline
\end{tabular}
}%
\end{table}

\begin{table}[ht]
\caption{Measurements parameters\label{tab:PIVflowdetails} }
\centering{
\begin{tabular}{l r}
\hline
Velocity target ($u^*$) &5m/s\\
Section area& 150x150 mm\\
Flow target &405 m$^3$/h\\
Re target &750000\\
Temperature start &19.5\textdegree\\
Temperature end &20\textdegree\\
N° of independent measurements &6\\\\
Mean flow measured: &404.2 m$^3$/h\\
\hline
\end{tabular}
}
\end{table}

\subsection{PIV recording parameters and processing}
The measurements have been carried out with HighSpeedStar camera system coupled with the software DaVis8 from LaVision. The light source used is a LDY300PIV laser model of Neodym:YLF family with characteristic wavelengths of $\lambda$= 1053nm and $\lambda$= 526nm. Laser time stabilizer and PTUX system have been utilized. This ensures all data to come from the same trigger, with 10ns of resolution, and accomplish synchronized data reading. The imaging technique used is the so called double-frame imaging, in which a series of coupled images is taken, with \textit{dt} time span between each frame, for the entire recording time. Details about the recording parameters are stated in Table \ref{tab:PIVdetails}. The camera position was perpendicular to the illumination plane. The particles used were hollow glass sphere, provided by LaVision, of a mean size of 9-13 $\mu$m and a density of 1.1 g/cm$^3$.

\begin{table}[ht]
\caption{PIV Recording parameters\label{tab:PIVdetails} }
\centering{
\begin{tabular}{l r}
\hline
Number of images&4367  \\
Recording time&1,8 s\\
Repetition rate&2441 Hz\\
\textit{dt} between coupled images&55 $\mu$s\\
\hline
\end{tabular}
}
\end{table}

The image processing is a fundamental step in order to obtain valid results. The concept at the base of image processing is the cross-correlation and it has been analyzed mathematically and already proven valid in past works \cite{westerweel_fundamentals_1997, westerweel_digital_1993}. The non-uniformity of light intensity captured in the image, due to multiple different source of discontinuity, can introduce noise which affect the correlation factor \cite{raffel_particle_2018}. In this experiment a multipass interrogation technique has been used with decreasing size from 94x94 px to 24x24 px with 50\% overlapping. The measurement uncertainty related to the velocity parameter is calculated from the correlation statistics \cite{wieneke_piv_2015}. PIV spatial calibration has been performed through a 3rd order polynomial interpolation and has given an RMS of fit of 0.441 px which, considering a camera resolution of 1.3 Megapixel, has been considered acceptable. The areas with higher uncertainty are in the early wake and related to the development of the Von Karman structures, in which the maximum estimated velocity uncertainty is 0.23 m/s.

\subsection{Proper Orthogonal Decomposition}
\label{PODmethod}
The POD is a mathematical procedure introduced in the 1960s by Lumley \cite{lumley_toward_1970} and further developed by Sirovich \cite{sirovich_turbulence_1987} in the 1980s. Its purpose is to represent turbulent fluid motion by partitioning it into distinct fluid modes, each characterized by specific kinetic energy. Through their superimposition, these modes collectively determine the overall turbulent kinetic energy (TKE) within the flow. This classification of fluid modes with different TKE levels enables the identification of which mode plays a predominant role in the specific turbulent flow motion. Additionally, it facilitates the identification of elusive modes that exist within the flow but are challenging to observe directly \cite{weiss_tutorial_2019}. These fluid modes are also referred to as \textit{coherent structures}, as defined by Fiedler \cite{fiedler_coherent_1988}, who describes them as "instability modes of the basic flow."

The POD method is particularly well-suited for processing PIV images due to its ability to handle a consistent a large structured dataset. In a PIV flow measurement, a set number of images is captured at a specific sampling rate, resulting in a defined time duration. Each image is a 2D matrix containing the instantaneous velocity vector \(\textit{V}\) for the given pixel resolution. If we define the matrix \(\overline{V}\) as the averaged velocity matrix over the measured time, meaning over the entire dataset, it becomes possible to derive the corresponding matrix of fluctuating velocity \(\textit{u'}\) for each instant. Such analysis enables the observation of flow dynamics by simply subtracting the time-averaged velocity matrix from the instantaneous velocity matrix. Considering \(\textit{Nt}\) as the number of pictures of dimensions \(\textit{w} \times \textit{h}\) pixels, every fluctuating velocity, in each position and at each time \(\textit{t}\), is reordered and included in the snapshot matrix \(U\):

\begin{gather}
    U=\begin{bmatrix} u_{11} & \hdots & u_{1n} \\ \vdots& &\vdots \\ u_{m1} & \hdots & u_{mn}
    \end{bmatrix}= \begin{bmatrix} u'(x_1,y_1,t_1)& \hdots & u'(x_w,y_h
,t_1) \\ \vdots& &\vdots \\ u'(x_1,y_1,t_m)& \hdots & u'(x_w,y_h
,t_m)
    \end{bmatrix}
\label{U}
\end{gather}

Where \textit{m} is the total number of images taken from PIV measurement. From this formula is now possible to extract the covariance matrix \textit{C}, which compute the statistical correlation between each point in the fluctuating velocity matrix dataset:
\begin{equation}
    C=\frac{1}{m-1}U^T U
    \label{C_dir}
\end{equation}

From linear algebra theory, the diagonal elements of the covariance matrix represent the variances of each specific pixel, or matrix point, while the non-diagonal terms represent the covariance between the different points. The dimension of this matrix is $(w \cdot h)^2$, which can result in a very computationally expensive calculation. 

As the covariance matrix is symmetric, from linear algebra is also possible to define it as:

\begin{equation}
    C=\Phi\Lambda\Phi^{-1}=\Phi\Lambda\Phi^T
    \label{C_dir_aut}
\end{equation}

where $\Phi$ and $\Lambda$ are eigenvector and eigenvalue of \textit{C}, and in explicit terms:

\begin{equation}
    C=\begin{bmatrix}
        \phi_{11} & \hdots &\phi_{1n}\\
        \vdots & & \vdots \\
        \phi_{m1} & \hdots &\phi_{mn}
    \end{bmatrix}
    \begin{bmatrix}
        \lambda_{1} & \hdots &0\\
        \vdots & & \vdots \\
        0 & \hdots &\lambda_{m}
    \end{bmatrix}
    \begin{bmatrix}
        \phi_{11} & \hdots &\phi_{n1}\\
        \vdots & & \vdots \\
        \phi_{1m} & \hdots &\phi_{mn}
    \end{bmatrix}
     \label{C_dir_exp}
\end{equation}

Where the eigenvalues are ordered from the largest to smallest. 

\begin{equation}
    \Phi=\begin{bmatrix}
        \phi_{11} & \hdots & \phi_{1n} \\ \vdots & & \vdots \\ \phi_{n1} & \hdots & \phi_{nn}
    \end{bmatrix}
     \label{Eigen_dir}
\end{equation}

Reordering $\Phi$ in the same fashion as the done for \textit{U} will allow for the determination of the so-called \textit{spatial modes}. So \textit{n} modes have been extracted, and to a higher eigenvalue correspond a higher contribution of the specific mode to the TKE. Is now possible to extract:

\begin{equation}
    A=U\Phi
\end{equation}

where \textit{A} represent the variance on each principal axis and is generally referred as time-coefficient matrix, of dimensions \textit{m} x \textit{n}, same as \textit{U},  while $\Phi$ is a \textit{n} x \textit{n}. This allow the extraction the specific spatial mode shape by reversing the formula:

\begin{equation}
    U=A\Phi^{-1}=A\Phi^T
\end{equation}

so that now:

\begin{equation*}
    U= \begin{bmatrix}
        a_{11} \\ \vdots \\ a_{m1}
    \end{bmatrix} 
    \begin{bmatrix}
        \phi_{11} & \hdots & \phi_{n1} 
    \end{bmatrix} + \hdots +
    \begin{bmatrix}
        a_{1n} \\ \vdots \\ a_{mn}
    \end{bmatrix} 
    \begin{bmatrix}
        \phi_{1n} & \hdots & \phi_{nn} 
    \end{bmatrix}\\
\end{equation*}

\begin{equation}
    =\begin{bmatrix}
        \tilde{u}^1_{11} & \hdots & \tilde{u}^1_{1n} \\
        \vdots & & \vdots\\
        \tilde{u}^1_{m1} & \hdots & \tilde{u}^1_{mn} \\
    \end{bmatrix} + \hdots +
    \begin{bmatrix}
        \tilde{u}^n_{11} & \hdots & \tilde{u}^n_{1n} \\
        \vdots & & \vdots\\
        \tilde{u}^n_{m1} & \hdots & \tilde{u}^n_{mn} \\
    \end{bmatrix}
\end{equation}

so that:
\begin{equation}
    U=\sum_{k=1}^{n} \tilde{U}^k
\end{equation}    

and the relationship of the specific mode TKE with the total TKE is defined as percentage from the following:
\begin{equation}
    TKE^k=\frac{\lambda^k}{\sum\lambda} 
\end{equation}

The method described above is generally referred as \textit{direct} method. A further development of POD algorithm has been done by Sirovich which has introduced the so-called \textit{Snapshot} method\cite{sirovich_turbulence_1987}, which is aimed to process larger dataset faster then the \textit{direct} method. The concept behind the \textit{Snapshot} method is to use the property of linear algebra inherent to transposition of matrices in order to compute smaller matrices, saving computational efforts. As seen in Formula \ref{U}, the \textit{U} matrix is a \textit{m} x \textit{n}. Building the canonical correlation matrix for \textit{U} will result in the correlation matrix \textit{C} as \textit{n} x \textit{n}, as in Formulas \ref{C_dir}-\ref{C_dir_exp}. In this modified approach, the new matrix\textit{$C_s$} is defined as:

\begin{equation}
    C_s=\frac{1}{m-1}UU^T
    \label{C_snap}
\end{equation}

Where \textit{m} represents the total number of PIV images taken during the measurement (\(N_t\)), resulting in an \(m \times m\) matrix. This is different from the previous method where the matrix was \(n \times n\), where \(n\) represented the number of data points in each image, equivalent to the product of the width (\(w\)) and height (\(h\)) of each image, in pixel. The procedure follows a similar path as before, involving the resolution of the eigenvalue problem. However, due to the inversion of the position of \(U\) in the calculation of the covariance matrix, \(C_s\) is constructed by averaging in space rather than time. Therefore, during the solution of the eigenvalue problem, instead of extracting the eigenvectors \(\Phi\), we obtain the so-called \textit{temporal modes} \(A_s\). However, the eigenvalues \(\lambda\) remain the same as those extracted from the previous method. Since the eigenvalues are consistent across both methods, the mode ordering remains unchanged. Thus, the mode that predominantly influences the total turbulent kinetic energy (TKE) will still be the first mode in both cases. 

\begin{equation}
    C_s=A_s\Lambda A_s^{-1}=A_s \Lambda A_s^T
    \label{C_snap_aut}
\end{equation}

After reordering the values from highest to lowest is possible to find:

\begin{equation}
    \Phi_s=U^TA_s
     \label{Eigen_snap}
\end{equation}

and the resulting mode \textit{k} can be extracted as:

\begin{equation}
    \tilde{U_s}^k =  A_s^k  \Phi_s^k = \begin{bmatrix}
        a_{s \hspace{1mm} 1k} \\ \vdots \\ a_{s \hspace{1mm} mk}
    \end{bmatrix}
    \begin{bmatrix}
        \phi_{s \hspace{1mm} 1k} & \hdots & \phi_{s \hspace{1mm} mk}
    \end{bmatrix}
\end{equation}

remembering that:

\begin{equation}
    U=\sum_{k=1}^{m} \tilde{U_s}^k
\end{equation}

\section{Numerical Setup}
\label{sec:numsetup}
\subsection{Computational domain}
Numerical model of the hydrofoil was created to investigate the vortex breakdown inside the test-section at different Reynolds numbers. The numerical investigation was conducted using a sequential approach to ensure proper verification and minimize unexpected error from the result initialization. Steady-state simulations based on the Shear Stress Transport RANS model were first employed to assess mesh adequacy and boundary conditions. Unsteady simulations using the Scale-Adaptive Simulation (SAS) model were then performed to capture large-scale wake unsteadiness \cite{arunn_master}. Finally, a high-resolution LES was carried out at the selected operating condition to resolve near-wall and wake dynamics. The computational domain consists of hydrofoil test-section along with the inlet and outlet conduit. The domain is divided into four subdomains: (1) inlet conduit with convergent section, $L=1.06$ m, (2) test-section with hydrofoil, $L=1.61$ m where chord ($c$) length of the hydrofoil is 0.25 m, (3) the downstream divergent section, $L=1.7$ m, (4) outlet conduit, $L=10$ m, to acquire development of the trailing edge vortex from the hydrofoil. Hexahedral mesh was created in the entire computational domain. All subdoimains are connected using unified mesh, and no interface between the subdoimains exists. The final mesh consists in a total of 500 million mesh nodes in the computational domain. The first node was placed at 0.5 micron and subsequent mesh expansion factor was 1.2 from the no-slip wall of the hydrofoil test-section.

\begin{figure}[ht]
\centering
\includegraphics[width=0.5\columnwidth]{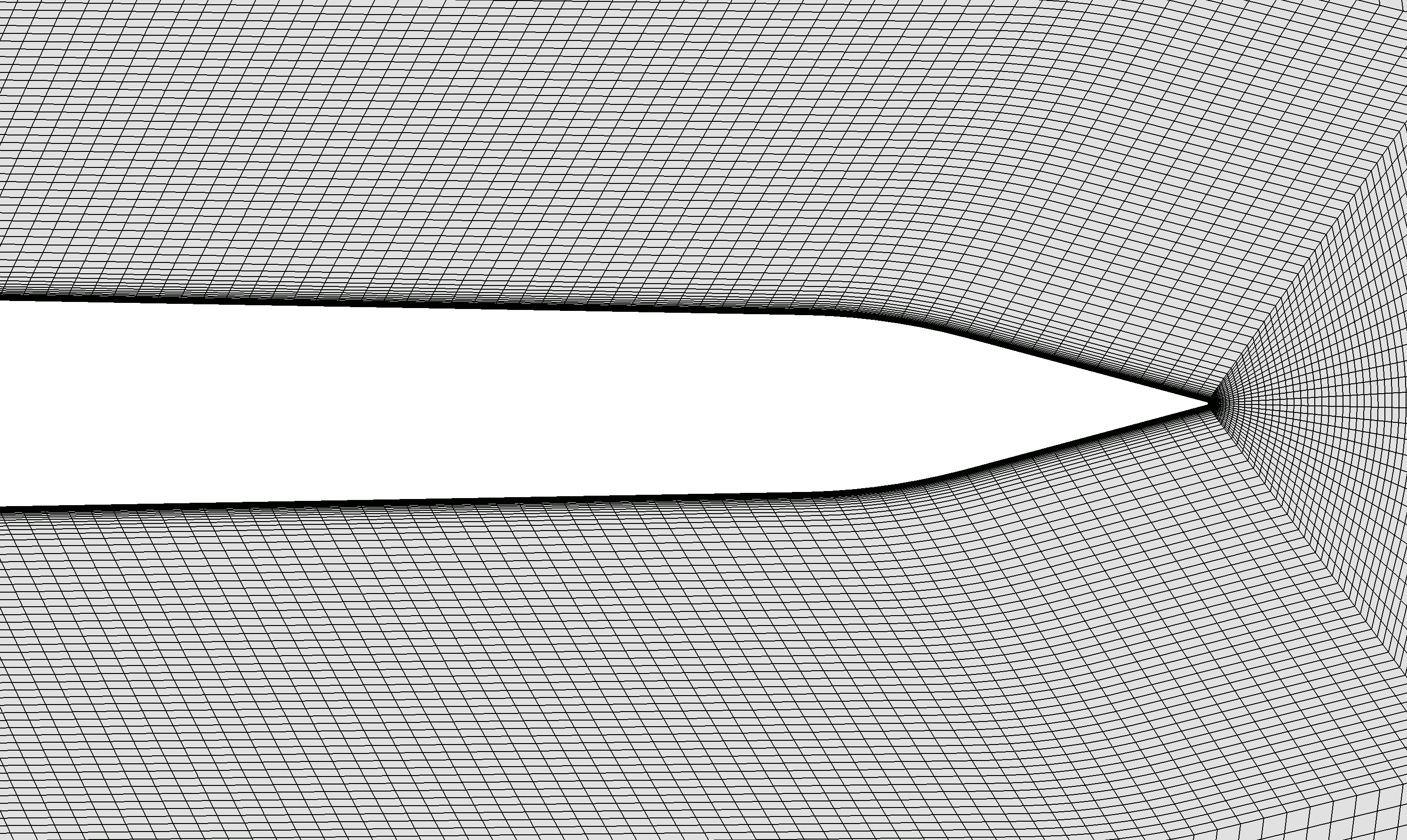}
\caption{Hexahedral mesh around hydrofoil (x-scale = 0.05 m).}
\label{fig:foilmeas}
\end{figure}

\begin{figure*}[ht]
\centering
\centering
\includegraphics[width=0.93\textwidth]{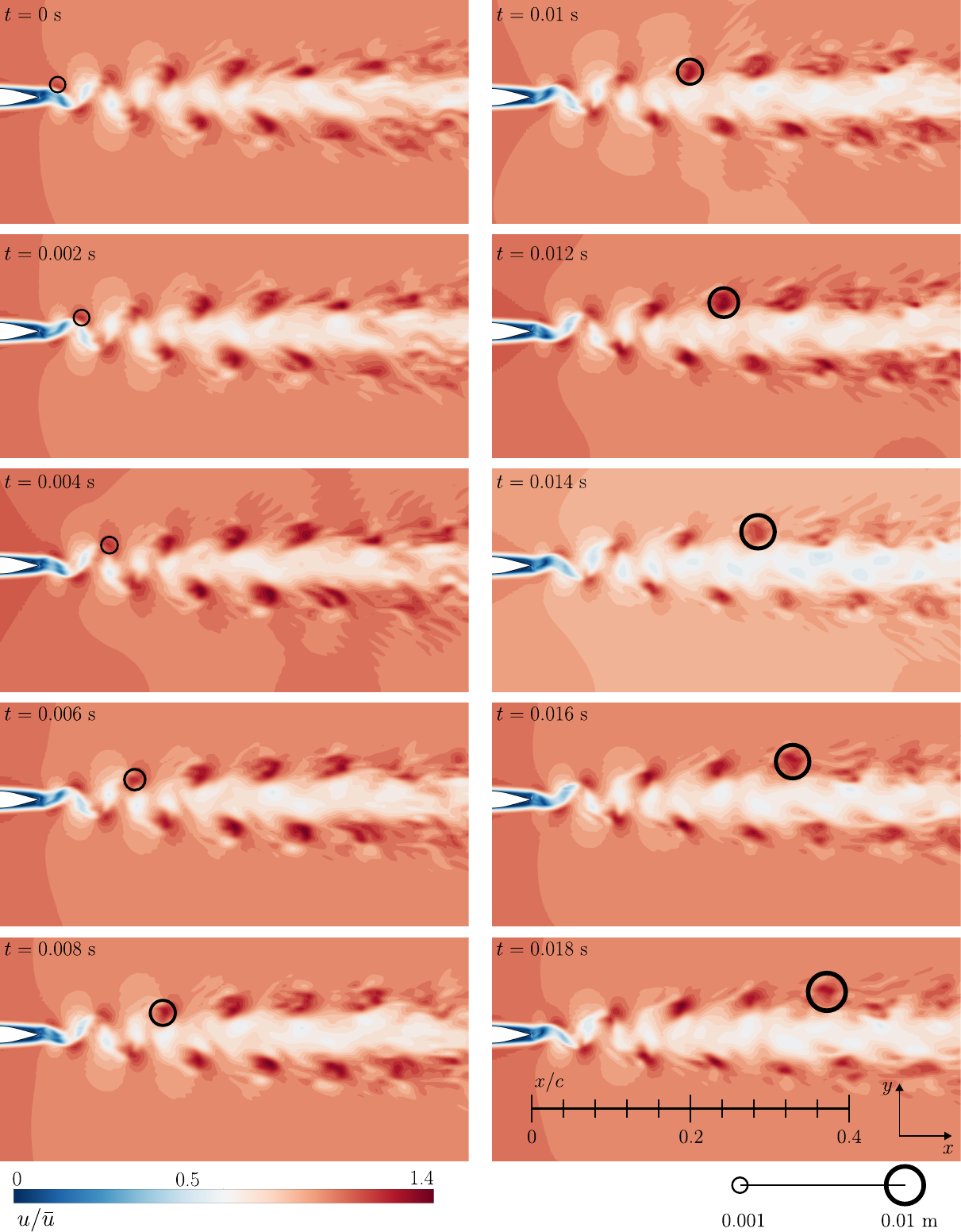}
\caption{Time dependent vortex breakdown at critical Re. \label{fig:timeDependent}}
\label{vortbreak}
\end{figure*}

\begin{figure*}[ht]
\centering
\includegraphics[width=1\textwidth]{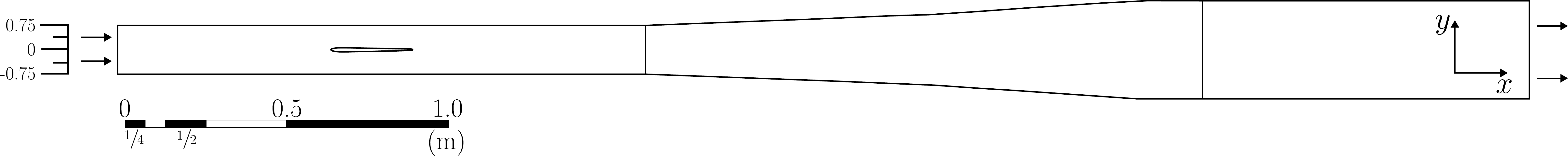}
\caption{Overview of the computational domain considered for the large eddy simulations. \label{fig:testSectionforLES}}
\end{figure*}

Steady state simulations have been carried out at beginning to study the overall performance of the numerical model, mesh and selected boundary conditions. Shear Stress Transport turbulence model with automatic wall function \cite{menter_review_2009} was considered for the simulations. Hexahedral mesh of 35 million nodes in all subdomains was created for the simulations. The maximum y$^+$ value was 5 in the computational domain. Four simulations at Re = 373 134, 746 268, 1 492 537 and 2 238 805 are carried out. The Reynolds numbers are computed using Equation ~\ref{Eq:Re}.

\begin{equation}
\label{Eq:Re}
\mathrm{Re}=\frac{\rho\cdot u^* \cdot D_h}{\mu}
\end{equation}

The hydraulic diameter ($D_h$) corresponds to the sides of the test-section, i.e., 0.15 m. The dynamic viscosity and density of the water at 20 C are 0.0010005 N-s/m$^2$, 1000 kg/m$^3$, respectively. The Reynolds number (Re$^*$ = 746 268) has been selected based on prior experimental characterization of the hydrofoil, corresponding to velocity $u^* = 5$ m/s. Free stream velocity is computed as, $u=Q/A$; where, $Q$ is the water flow rate in m\textsuperscript{3} s\textsuperscript{-1}, $A$ is the cross sectional area of the hydrofoil test section in m\textsuperscript{2}.

The numerical results of steady state simulations converged well (the maximum residual for mass, momentum and turbulent quantities was 0.0001) and showed expected velocity and pressure field in the computational domain. For the next step of verification, unsteady simulations using Scale Adaptive Simulation (SAS) approach are carried out. SAS modelling approach is considered to study vortex breakdown from the trailing edge. SAS model has demonstrated its capability to capture unsteady vortex breakdown \cite{menter_scale-adaptive_2010,egorov_scale-adaptive_2010}.

The locations selected for extracting velocity values along the normal ($y$) axis are shown in Figure \ref{fig:velprofileloc}. The reference location corresponds to the trailing edge, where $x=0$ m and $y=0$ m. The velocity values have been extracted along the lines also at Reynolds numbers of 373 134, 746 268, 1 492 537 and 2 238 805 (see Appendix \ref{APP:LES}). The normalized velocity values for $p5$, $p6$, $p7$ are presented in Figure  ~\ref{fig:velprofilecomp}, where o  n x-axis, $U/U\textsubscript{bulk}$ is the normalized velocity, where $U\textsubscript{bulk}$ is the bulk velocity at the Reynolds number corresponding to a velocity of 5 m/s. The velocity values are extracted for all profile lines at final time stamp of the of the numerical results. However, we presented only for three lines for clear visualization. There is no significant difference in velocity for the other lines and overall trend is similar. A steep gradient in the wake attached to the trailing edge can be seen, due to wall shear and flow separation from the hydrofoil boundary later. The wake effect diminishes at far away from the training edge and the velocity gradient is almost uniform.

The final simulations were conducted using LES approach at  Re$^*$ = 746 268. The mesh in the hydrofoil section was refined further from 35 to 500 million nodes. The simulations are performed using supercomputer with 144 cpu processors and the total simulation time with the supercomputer was 30 days. The LES was initialized using results of SAS modelling approach at the same Reynolds number. The computational domain was reduced to the hydrofoil test section and downstream divergent section, see Figure  ~\ref{fig:testSectionforLES}. Hence, the inlet and outlet boundary conditions corresponds to the inlet of the hydrofoil test section and outlet of the divergent section. The required pressure and velocity boundary data at the exact locations were acquired from the SAS simulation results.

\section{Results and discussion}
\subsection{Wake flow characteristics}
From the given PIV field of view, measuring 56.85×45 mm and located 13.15 mm from the trailing edge, velocity measurements were taken at cross-sections positioned 25 mm, 35 mm, and 55 mm downstream of the trailing edge (section \textit{$p_{5}$}, \textit{$p_{6}$} and \textit{p\textsubscript{7}} in Figure \ref{fig:velprofileloc}).

 \begin{figure*}[ht]
    \centering
    \begin{minipage}{0.475\textwidth} 
\centering
\includegraphics[width=1\columnwidth]{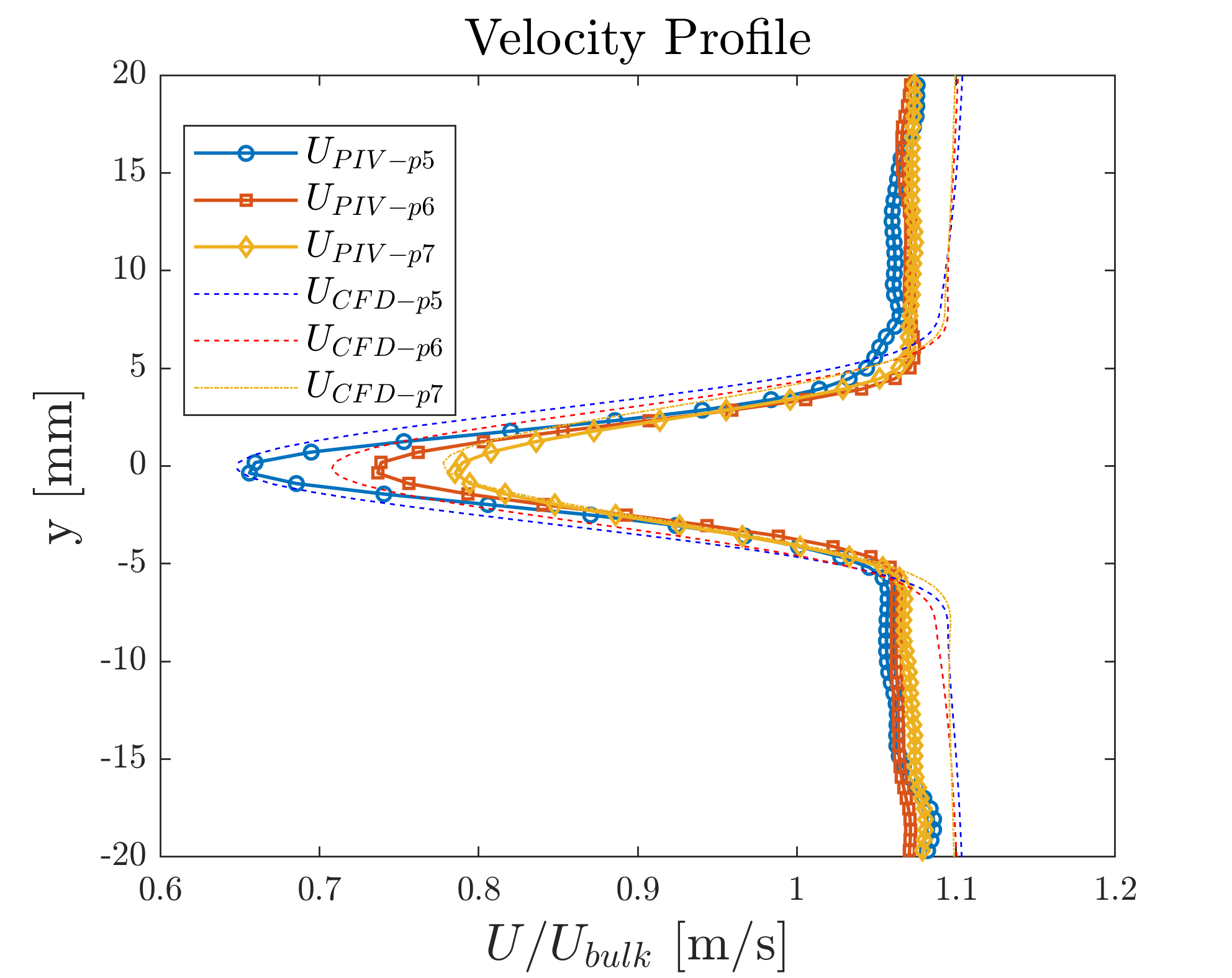}
\caption{Normalized velocity profiles at the 3 different sections considered from the trailing edge, for both LES and PIV.
\label{fig:velprofilecomp}}
    \end{minipage}\hfill
    \begin{minipage}{0.475\textwidth} 
\centering
\includegraphics[width=1\columnwidth]{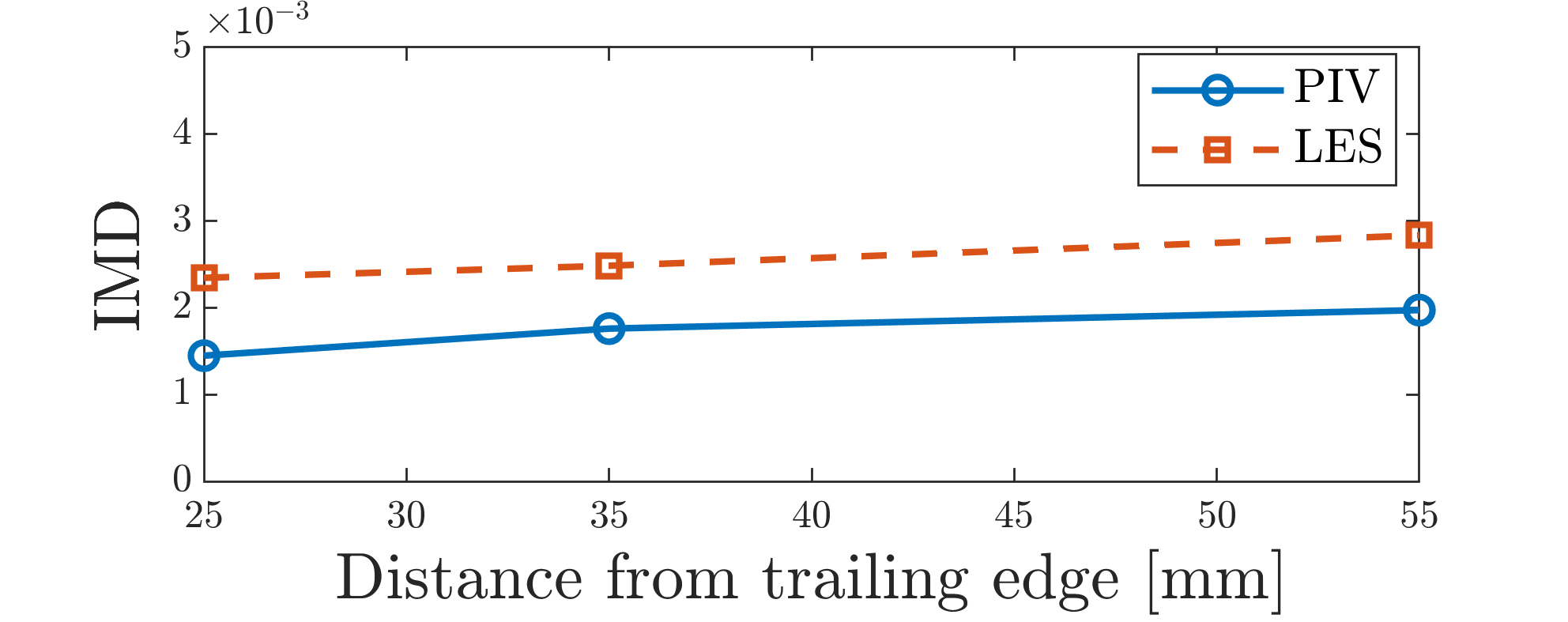}
\caption{Integrated Momentum Deficit for PIV and LES at the cross sections \textit{p5, p6, p7}.
\label{fig:IMD}}
    \end{minipage}
 \end{figure*}

As expected, the normalized velocity profiles show a velocity reduction in the central region of the test section near the trailing edge, while gradually moving towards the bulk value further downstream and transversely, where they stabilize between 1.075-1.1 the bulk velocity, depending on the dataset. The velocity profile obtained from the PIV measurements is based on time averaging over the duration of each individual measurements, while the corresponding LES velocity profile is derived from samples collected over approximately 3000 time steps. The comparison between the two velocity results has been done in terms of RMS metric and, in order to better compare the velocity results obtained by both PIV and LES simulation, Integrated Momentum Deficit (IMD). The RMS and IMD results for the three considered cross-sections are summarized in Table \ref{tab:RMS_IMD} and Figure \ref{fig:IMD}. The RMS results indicate good agreement between the two datasets, with RMS values decreasing as the distance from the trailing edge increases. This trend is expected, as the flow transitions away from the trailing-edge region toward the bulk velocity. Regarding the IMD, it is calculated as:

\begin{equation}
    IMD=\int (1 - \bar{u}) \, \bar{u} \, dy
\end{equation}

where $\bar{u} = {u}/{U_\infty}$ is the normalized velocity. This parameter quantifies the overall momentum loss in the wake through the integration of the normalized velocity deficit across the measurement span. 

\begin{table}[ht]
\caption{CFD-PIV RMS of the velocity results and Integrated Momentum Deficit (IMD) \label{tab:RMS_IMD} }
\centering{
\begin{tabular}{l c c c}
\hline
Section & RMS & IMD\textsubscript{PIV} & IMD\textsubscript{LES} \\
\hline
\textit{p5}& 3.92 \% & 1.44e-03 & 2.34e-03\\
\textit{p6}& 3.56 \% & 1.76e-03 & 2.48e-03\\
\textit{p7}& 2.38 \% & 1.97e-03 & 2.83e-03\\
\hline
\end{tabular}
}
\end{table}

The IMD results indicates that the values obtained from LES are consistently slightly higher than those derived from PIV at all downstream locations. This behavior is expected, primarily due to the limited transversal extent of the PIV measurement domain, which does not capture the outer regions of the wake where the velocity deficit, although small, still contributes to the total integrated momentum loss. In contrast, the LES data include the entire wake cross-section and thus account for these outer regions together with the resolution of finer flow structures. Additionally, differences in temporal sampling contribute to the observed discrepancy: the LES velocity fields are averaged over a substantially larger number of time steps, which may capture more persistent low-velocity regions, whereas the PIV measurements are based on time averaging over shorter acquisition intervals.

\begin{figure}[ht]
\centering
\includegraphics[width=0.5\columnwidth]{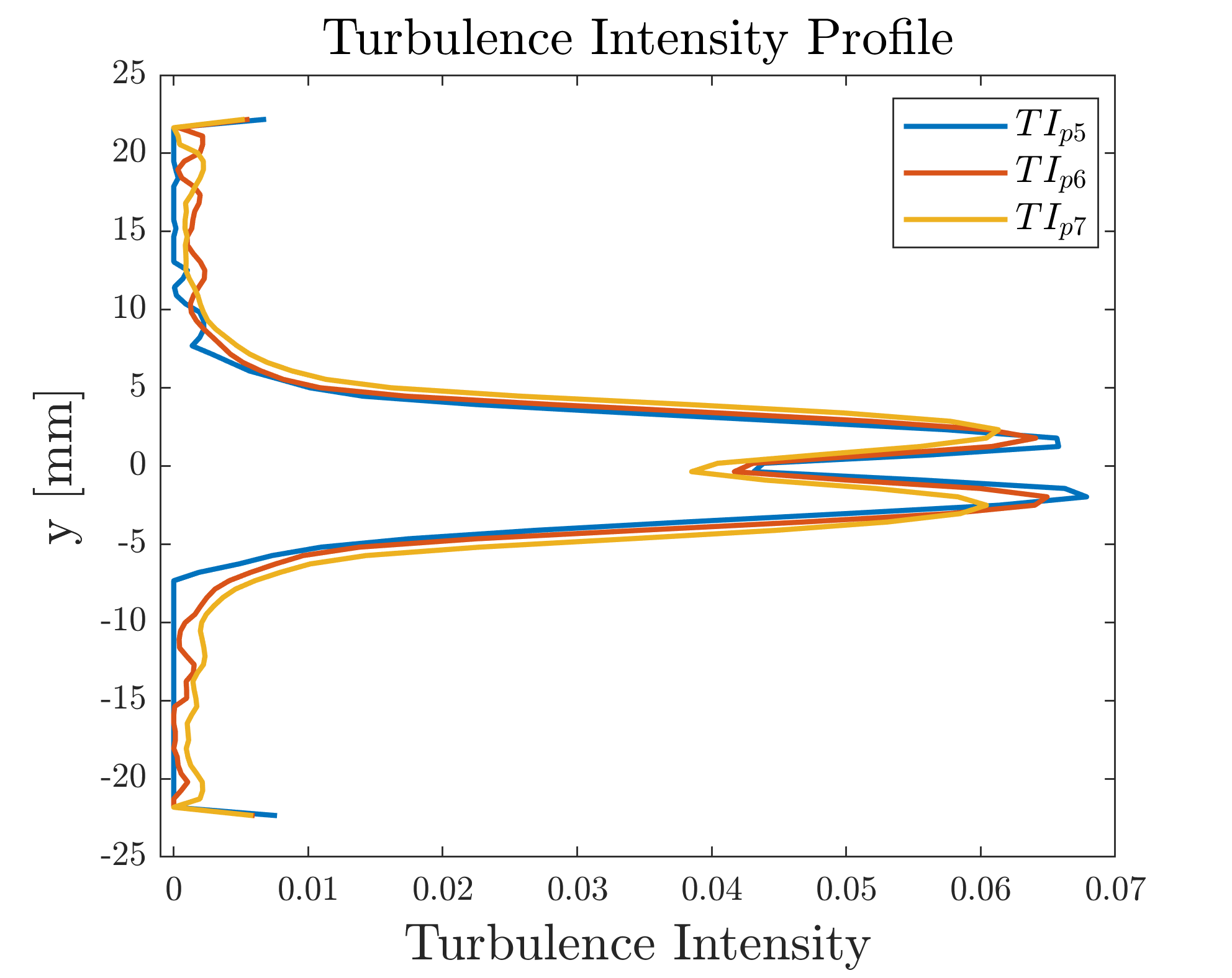}
\caption{Turbulence intensity profiles for the streamwise velocity component of PIV measurement at the 3 different cross sections considered. \label{fig:turbulenceprofiles}}
\end{figure}

The turbulent intensity profiles of the PIV dataset, measured in the three considered cross-sections, shows an increase corresponding to the two location close to the centerline where the Von Karman structures develops, while the interaction between vortices edges generates the value corresponding to the centerline, which is smaller than the one corresponding to the vortices path. Figure~\ref{fig:y_compSpectra} presents the power spectral density of the vertical cross stream velocity component sampled in the wake at $x/d =$ 7.5, $y/h =$ 0. The peak of the power spectrum has been found to be close to 295 Hz. This peak is commonly associated with the dominant vortex shedding frequency \cite{sagmo_piv_2019,fernandez-aldama_characterization_2025} and, accordingly, this frequency also basically corresponds to the dominant transverse coupled mode identified by the POD analysis and discussed in the following paragraph.

\subsection{Proper orthogonal decomposition of wake}

\begin{figure*}[ht]
    \centering
    \begin{minipage}{0.475\textwidth} 
\centering
\includegraphics[width=1\columnwidth]{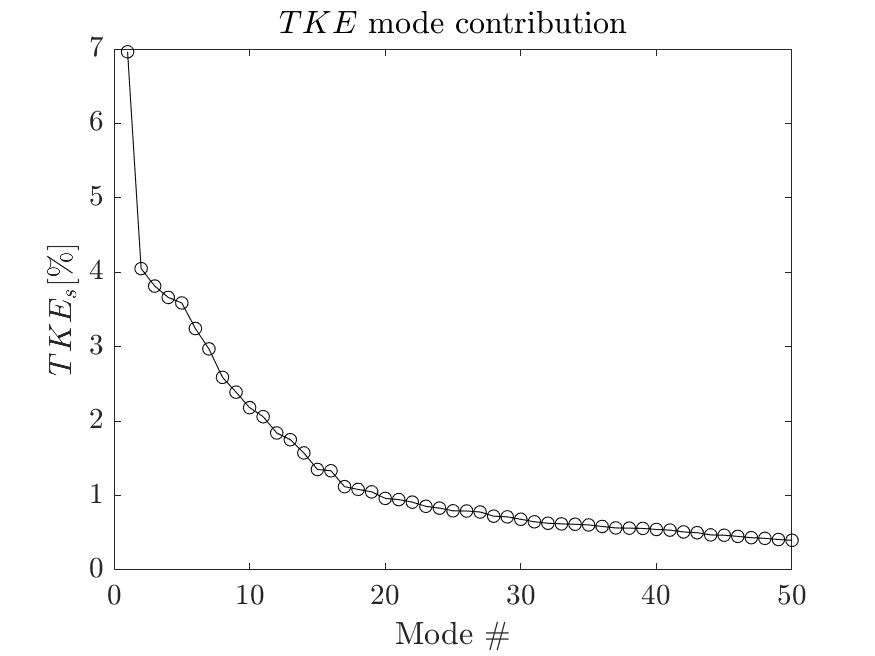}
\caption{TKE contribution of singular modes. }
\label{fig:TKEcontribution}
    \end{minipage}\hfill
    \begin{minipage}{0.475\textwidth} 
\centering
\includegraphics[width=1\columnwidth]{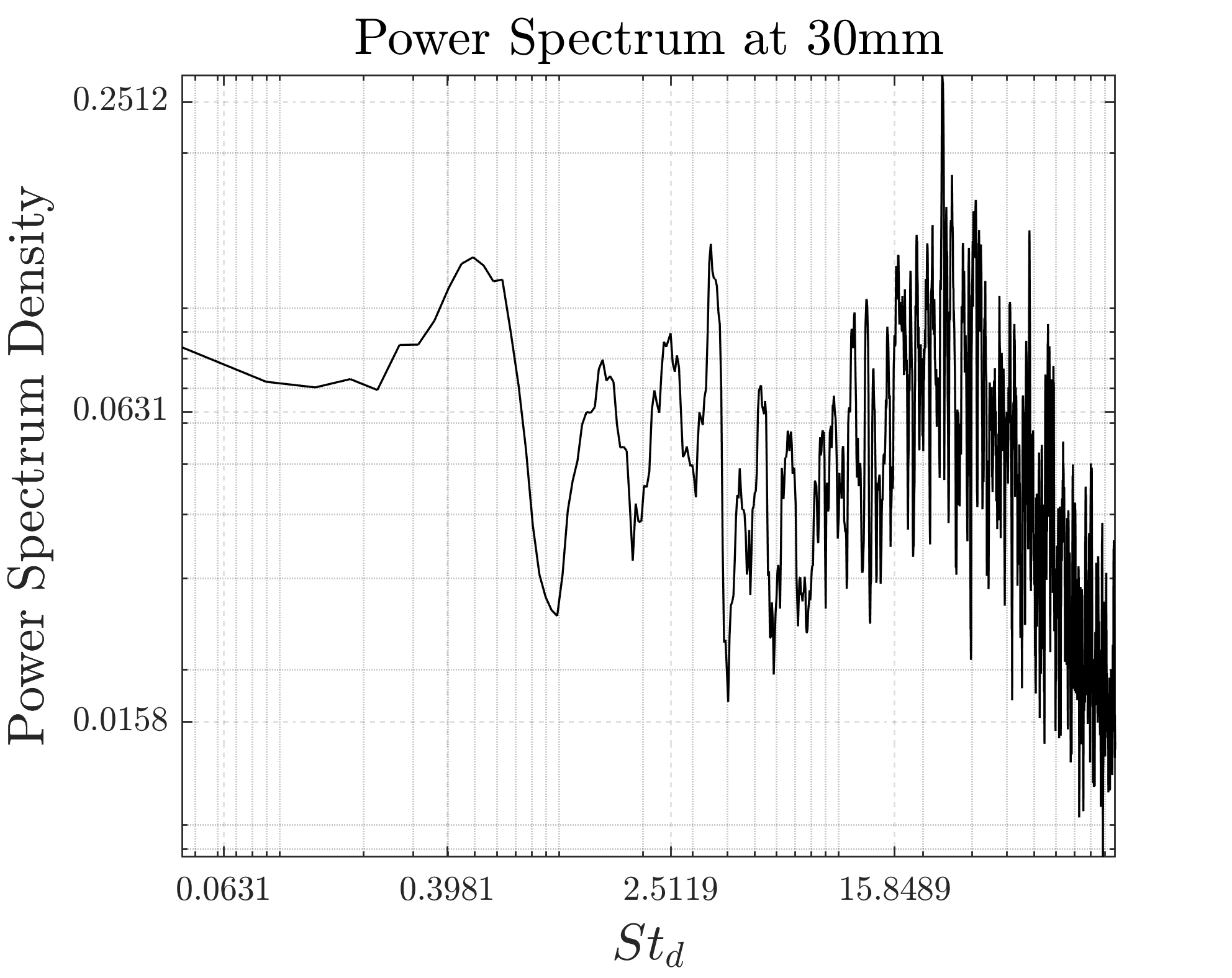}
\caption{Power spectral density of the vertical cross stream velocity component of PIV measurement, sampled in the wake at $x/d =$ 7.5, $y/h =$ 0. \label{fig:y_compSpectra}}
    \end{minipage}
 \end{figure*}

To understand the behavior and development of coherent structures on a large scale, a POD analysis of the PIV measurement set has been conducted. Following the POD Snapshot method as described in Section \ref{PODmethod} \cite{sirovich_turbulence_1987}, an analysis of the singular mode contribution over the total TKE present in the flow has been done for the first 50 modes, which together contribute to 67.7\% of the total TKE, as can be seen in Figure  \ref{fig:TKEcontribution}.

The absence of modes with high energy relative to the total TKE along with the overall low percentage of total TKE represented by the first 50 modes indicates that the setup, defined by the combination of flow and hydrofoil characteristics, results in low-energy wake velocity fluctuations that are well distributed across the modes. The likely cause of these low-energy wake structures is the slender profile of the hydrofoil, positioned at a 0° angle of attack relative to the flow, combined with its sharp trailing edge and the high flow velocity (5 m/s). It is noted that POD provides an energy-optimal, statistical decomposition of the flow field, therefore the extracted modes are ranked by energetic content and should not be interpreted as dynamically independent entities.

The first 3 modes are the most energetic modes and better represent the flow characteristics. 

\begin{table}[ht]
 \caption{TKE \% of the first 3 POD modes\label{tab:TKEfraction} }
 \centering{
 \begin{tabular}{c c }
    \hline
   Mode 1& 6.96 \% \\
   Mode 2& 4.05 \% \\
   Mode 3& 3.81 \% \\
   \hline
   \end{tabular}
   }
 \end{table}

 \begin{figure*}[ht]
    \centering
    \begin{minipage}{0.475\textwidth} 
        \centering
        \includegraphics[width=1\linewidth]{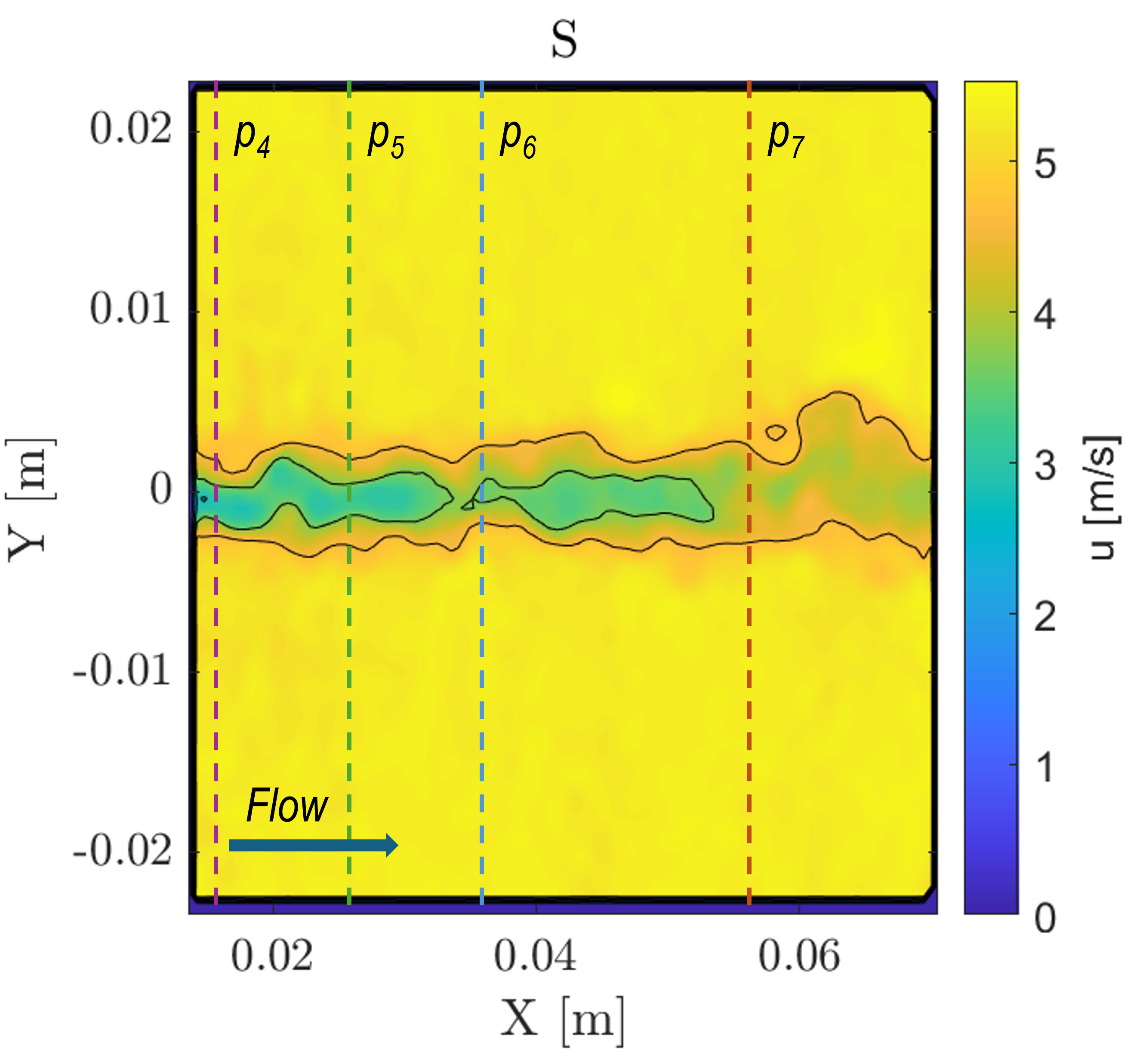} 
        \caption{Velocity field measured from PIV in a random instant.}
        \label{S}
    \end{minipage}\hfill
    \begin{minipage}{0.475\textwidth} 
        \centering
        \includegraphics[width=\linewidth]{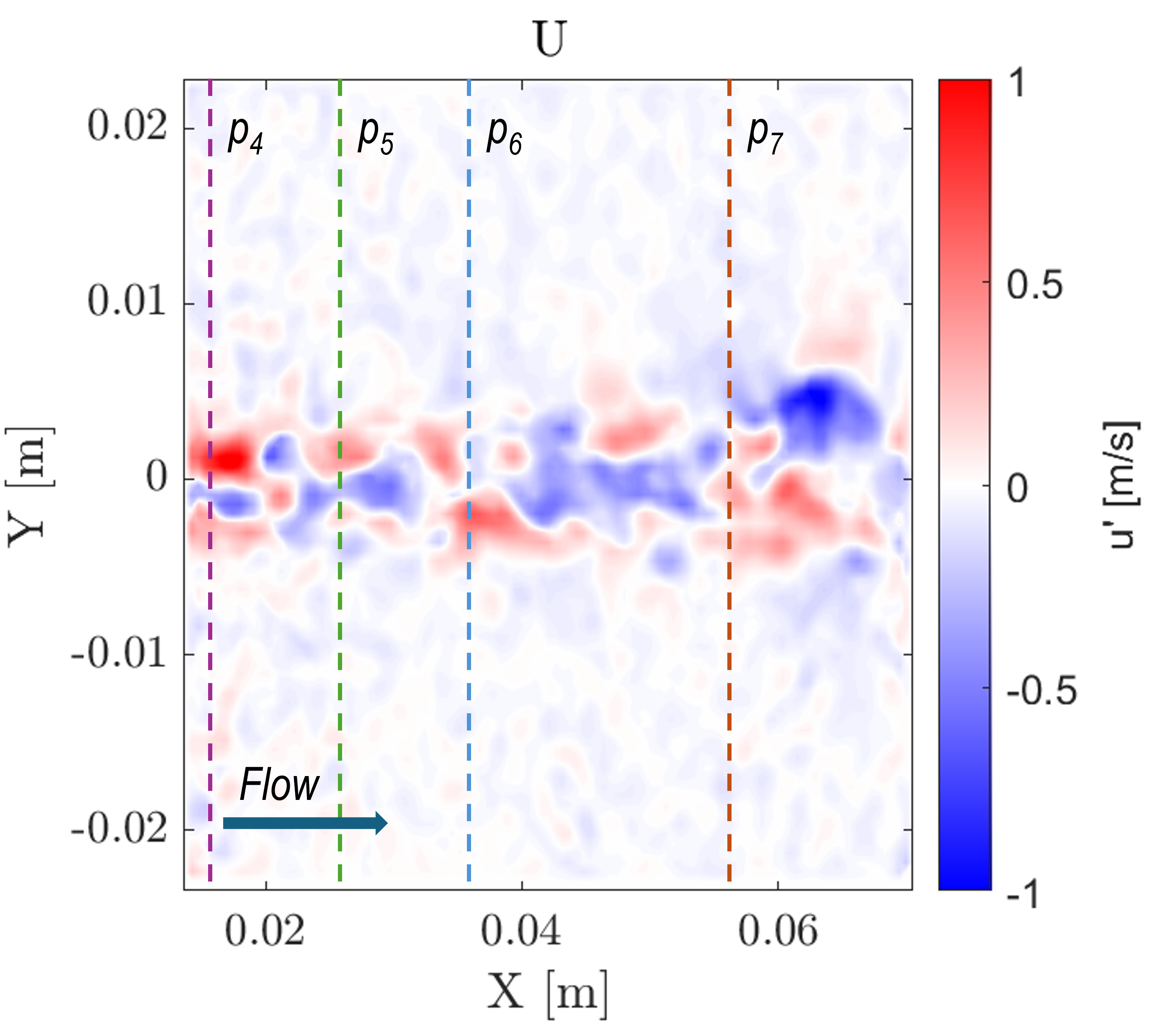} 
        \caption{Fluctuating velocity field related to matrix \textit{S} from Figure  \ref{S}. }
        \label{U}
    \end{minipage}
 \end{figure*}

\begin{figure*}[ht]
    \centering
    \begin{minipage}{0.5\textwidth} 
        \centering
        \includegraphics[width=1\linewidth]{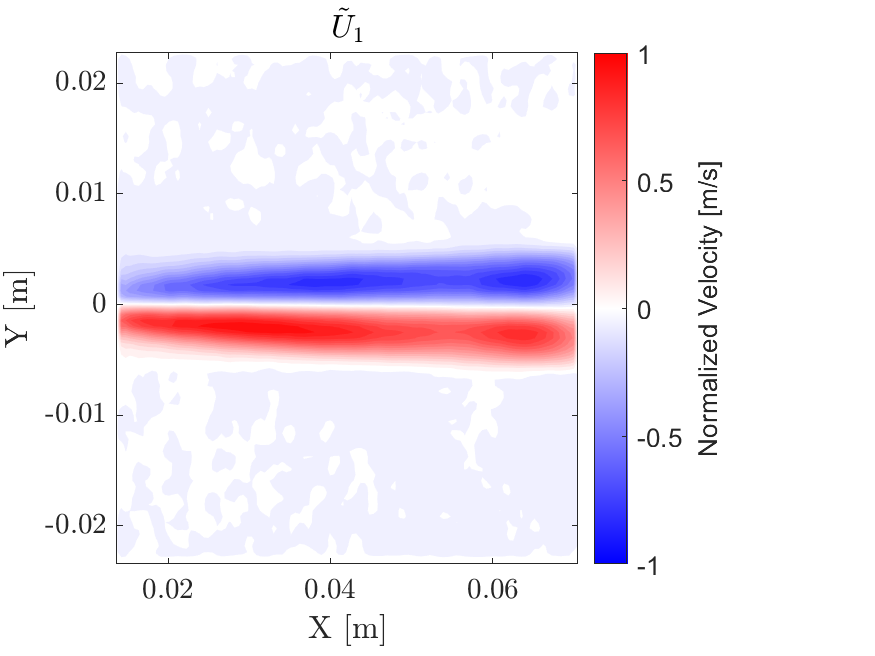} 
        \caption{POD mode 1.}
        \label{mode1}
    \end{minipage}\hfill
    \begin{minipage}{0.475\textwidth} 
        \centering
        \includegraphics[width=\linewidth]{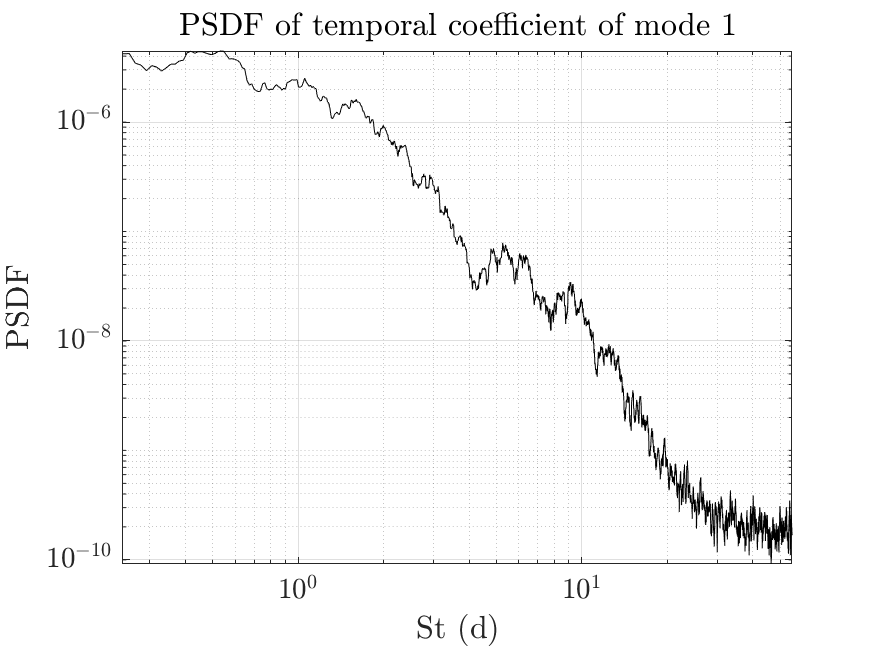} 
        \caption{Power spectrum density function of POD mode 1.}
        \label{psdf1}
    \end{minipage}
 \end{figure*}

 Their overall contribution towards the total fluctuating velocity can be assessed by comparing the velocity fraction associated with the first three modes to the overall fluctuating velocity, showing that the contribution of the three dominant modes does not represent a consistent fraction of the total velocity fluctuations. In particular, modes 2 and 3 exhibit a predominantly oscillatory behavior that does not consistently align with the evolution of the total fluctuating velocity. By contrast, mode 1 more closely follows the trend of the overall fluctuations, indicating that it captures the primary temporal dynamics of the fluctuating velocity, while the lower-raked modes mainly contribute through periodic variations rather than through a sustained share of the total fluctuation.

  \begin{figure*}[ht]
    \centering
    \begin{minipage}{0.5\textwidth} 
        \centering
        \includegraphics[width=1\linewidth]{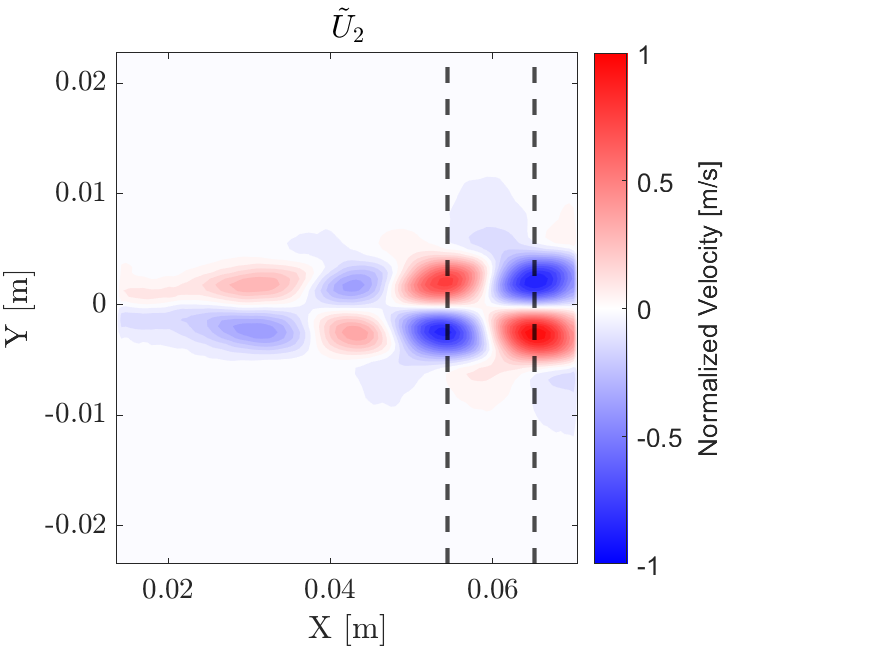} 
        \caption{POD mode 2.}
        \label{mode2}
    \end{minipage}\hfill
    \begin{minipage}{0.5\textwidth} 
        \centering
        \includegraphics[width=\linewidth]{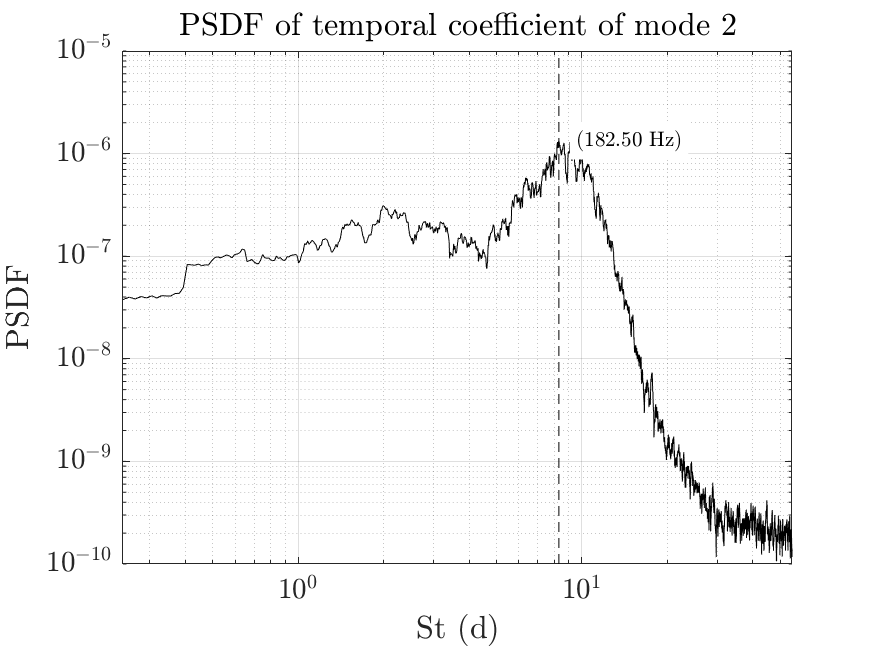} 
        \caption{Power spectrum density function of POD mode 2.}
        \label{psdf2}
    \end{minipage}
    \\
    \begin{minipage}{0.5\textwidth} 
        \centering
        \includegraphics[width=1\linewidth]{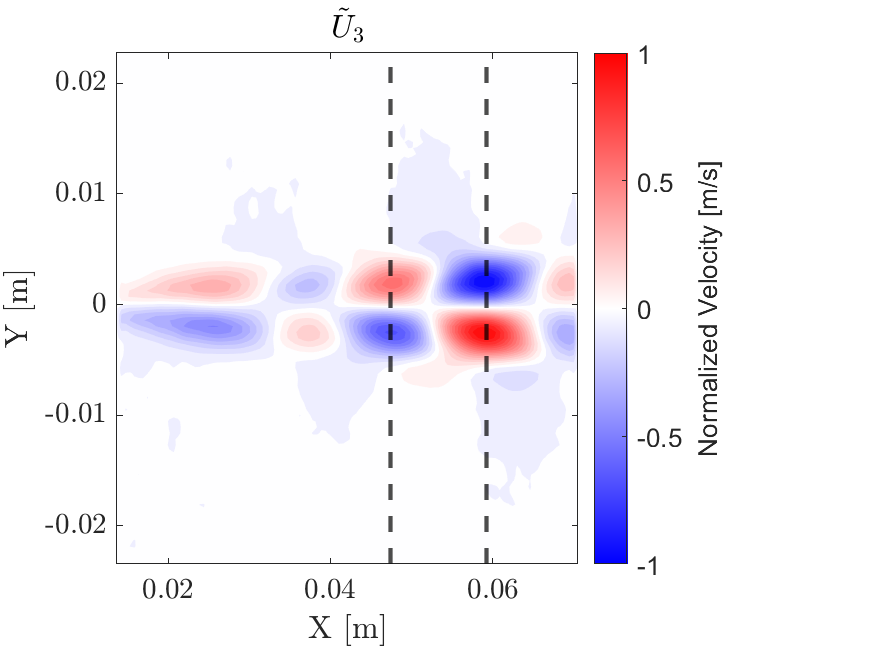} 
        \caption{POD mode 3.}
        \label{mode3}
    \end{minipage}\hfill
    \begin{minipage}{0.5\textwidth} 
        \centering
        \includegraphics[width=\linewidth]{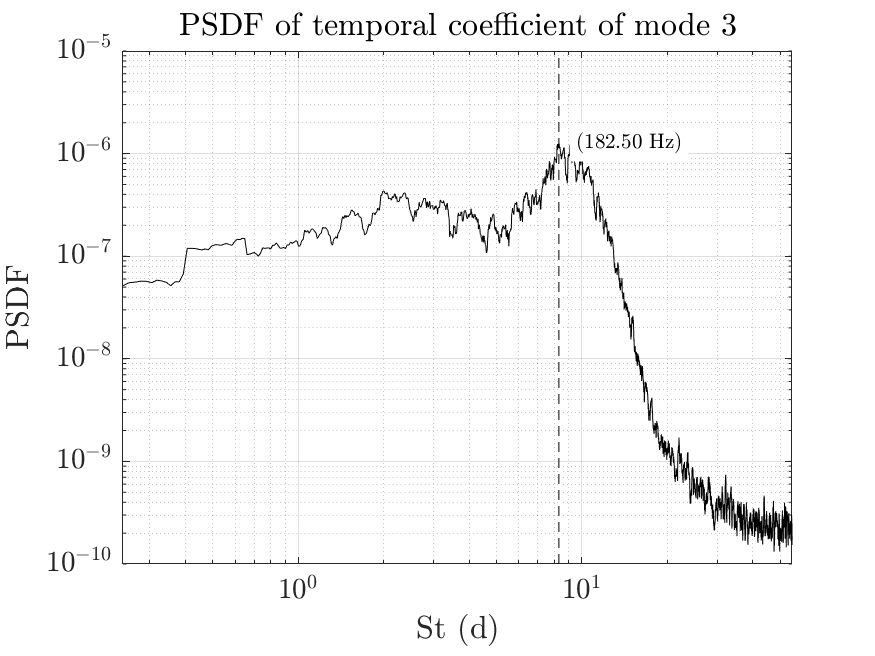} 
        \caption{Power spectrum density function of POD mode 3.}
        \label{psdf3}
    \end{minipage}\\
        \centering
    \begin{minipage}{0.49\textwidth} 
        \centering
        \includegraphics[width=1\linewidth]{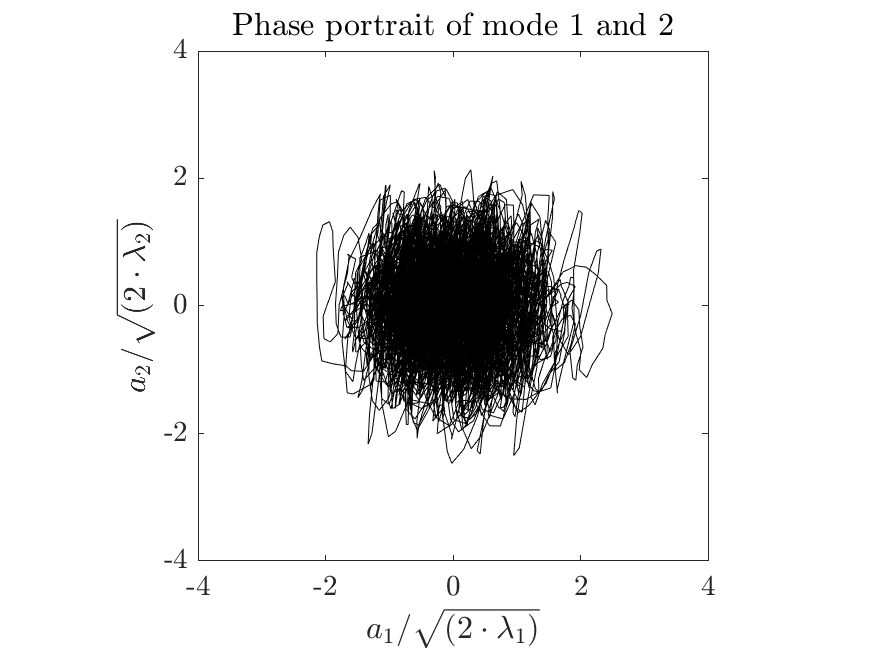} 
        \caption{Phase portrait of temporal coefficient of modes 1 and 2.}
        \label{phase12}
    \end{minipage}\hfill
    \begin{minipage}{0.49\textwidth} 
        \centering
        \includegraphics[width=1\linewidth]{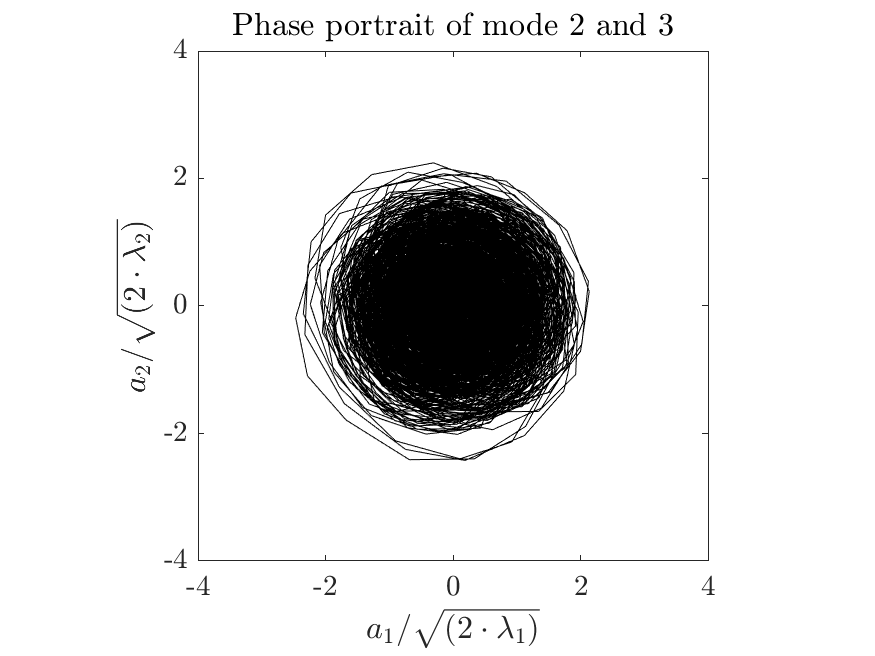} 
        \caption{Phase portrait of temporal coefficient of modes 2 and 3.}
        \label{phase23}
    \end{minipage}
\end{figure*}

\begin{figure*}[ht]
    \centering
    \begin{minipage}{0.5\textwidth} 
        \centering
        \includegraphics[width=1\linewidth]{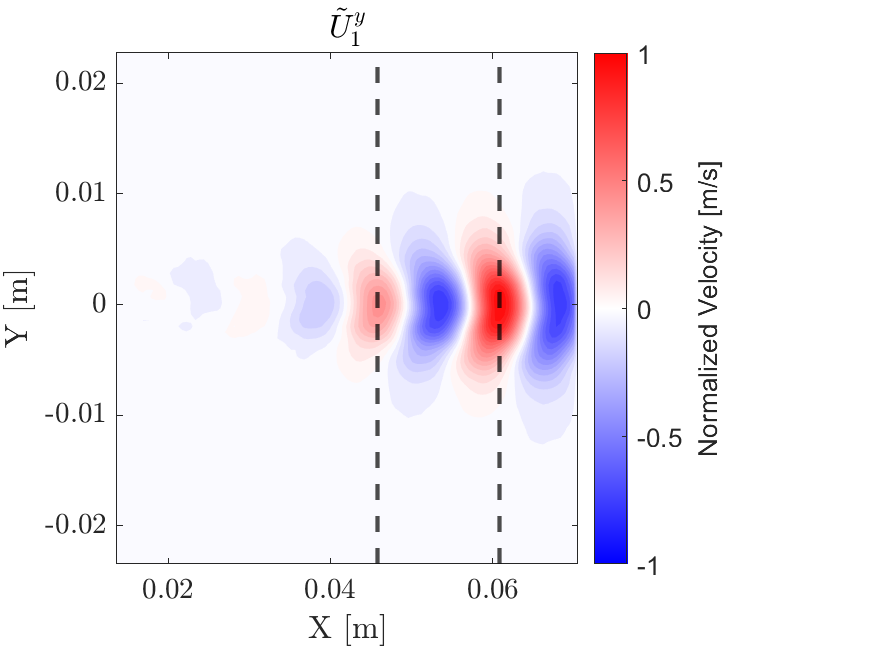} 
        \caption{POD mode 1, transversal component.}
        \label{mode1y}
    \end{minipage}\hfill
    \begin{minipage}{0.5\textwidth} 
        \centering
        \includegraphics[width=1\linewidth]{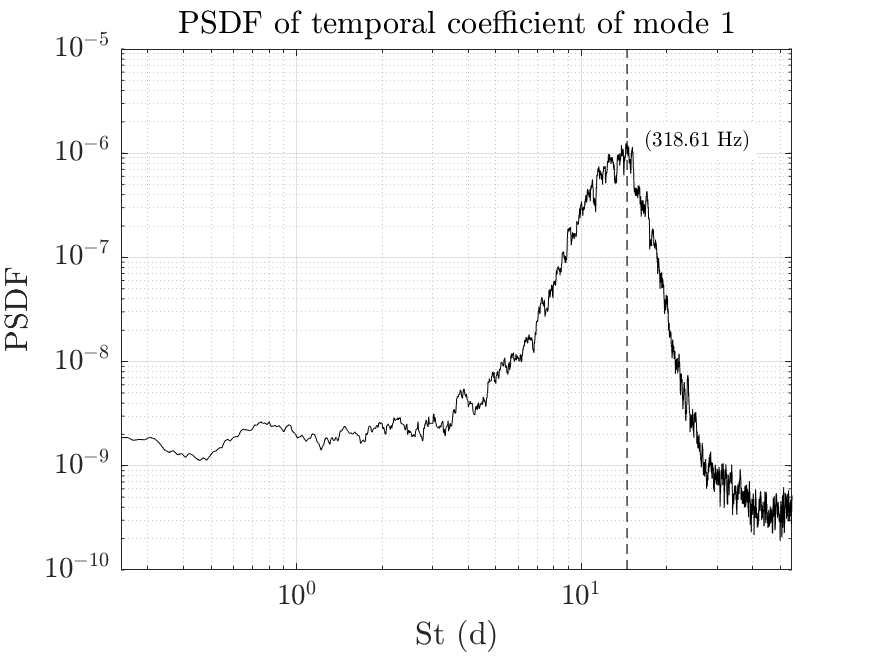} 
        \caption{Power spectrum density function of transverse mode 1.}
        \label{psdfmode1y}
    \end{minipage}
    \\
        \centering
    \begin{minipage}{0.5\textwidth} 
        \centering
        \includegraphics[width=1\linewidth]{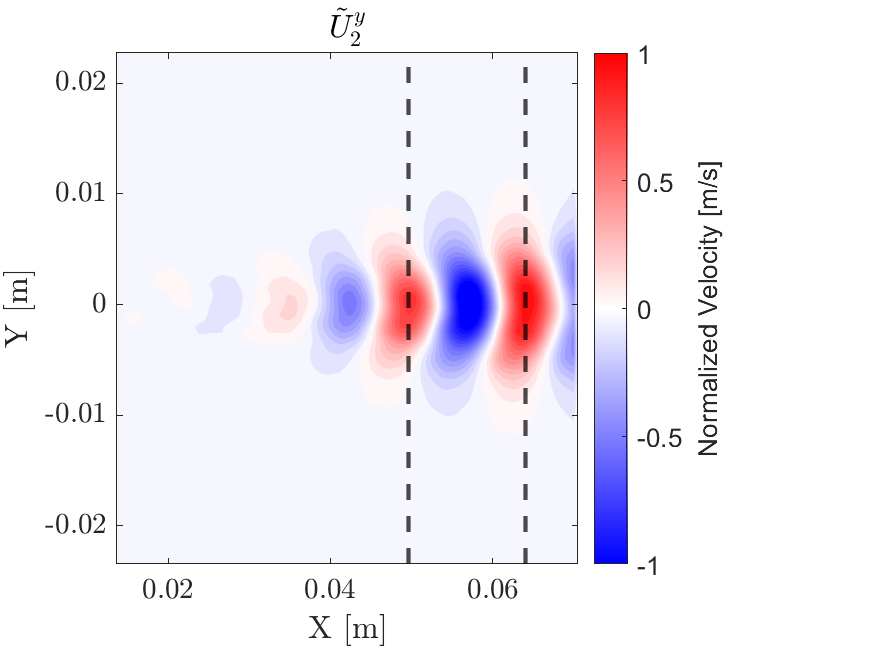} 
        \caption{POD mode 2, transversal component.}
        \label{mode2y}
    \end{minipage}\hfill
    \begin{minipage}{0.5\textwidth} 
        \centering
        \includegraphics[width=1\linewidth]{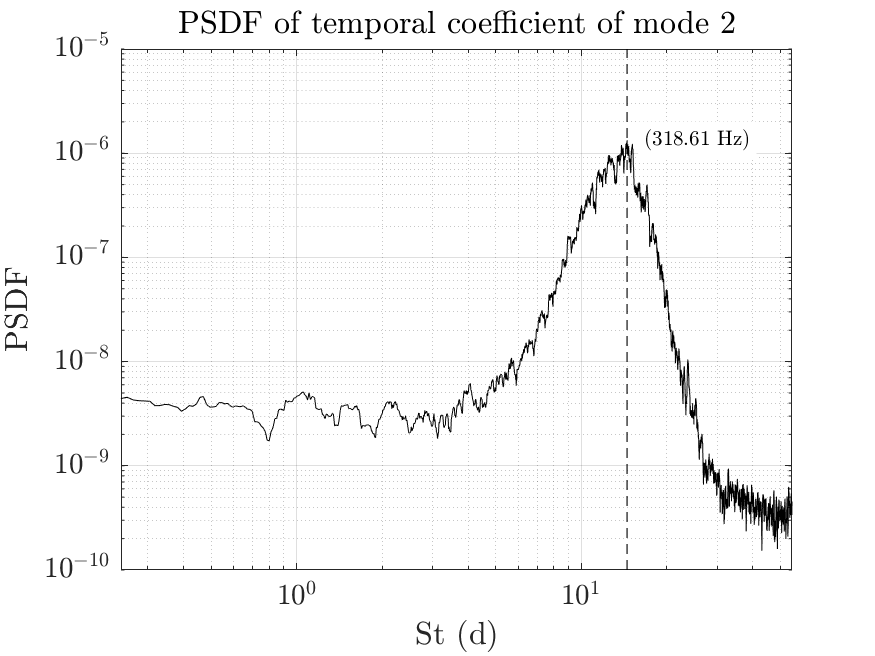} 
        \caption{Power spectrum density function of transverse mode 2.}
        \label{psdfmode2y}
    \end{minipage}
    \\
        \centering
    \begin{minipage}{0.5\textwidth} 
        \centering
        \includegraphics[width=1\linewidth]{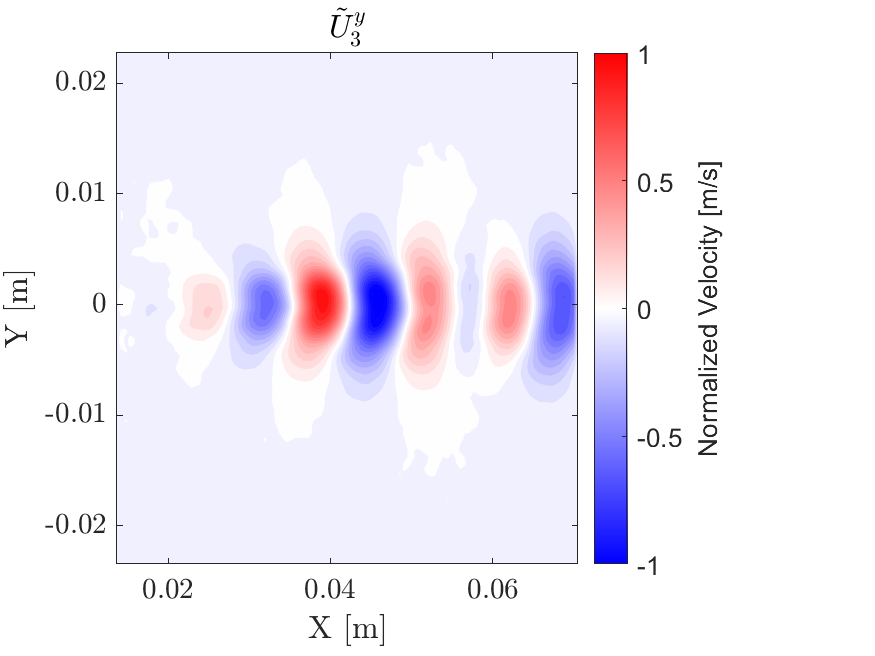} 
        \caption{POD mode 3, transversal component.}
        \label{mode3y}
    \end{minipage}\hfill
    \begin{minipage}{0.5\textwidth} 
        \centering
        \includegraphics[width=1\linewidth]{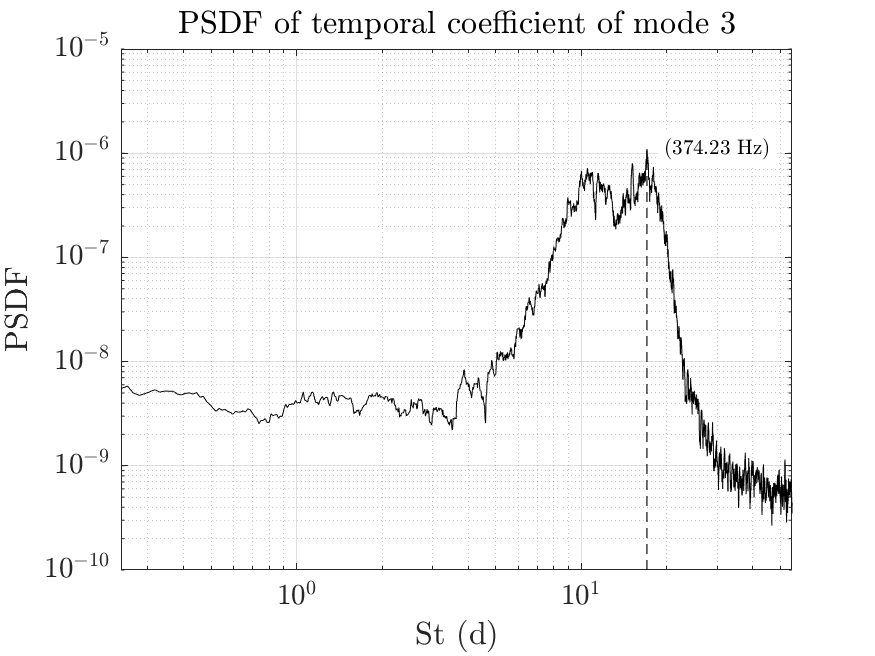} 
        \caption{Power spectrum density function of transverse mode 3.}
        \label{psdfmode3y}
    \end{minipage}
\end{figure*}

\begin{figure*}[ht]
\includegraphics[width=1\textwidth]{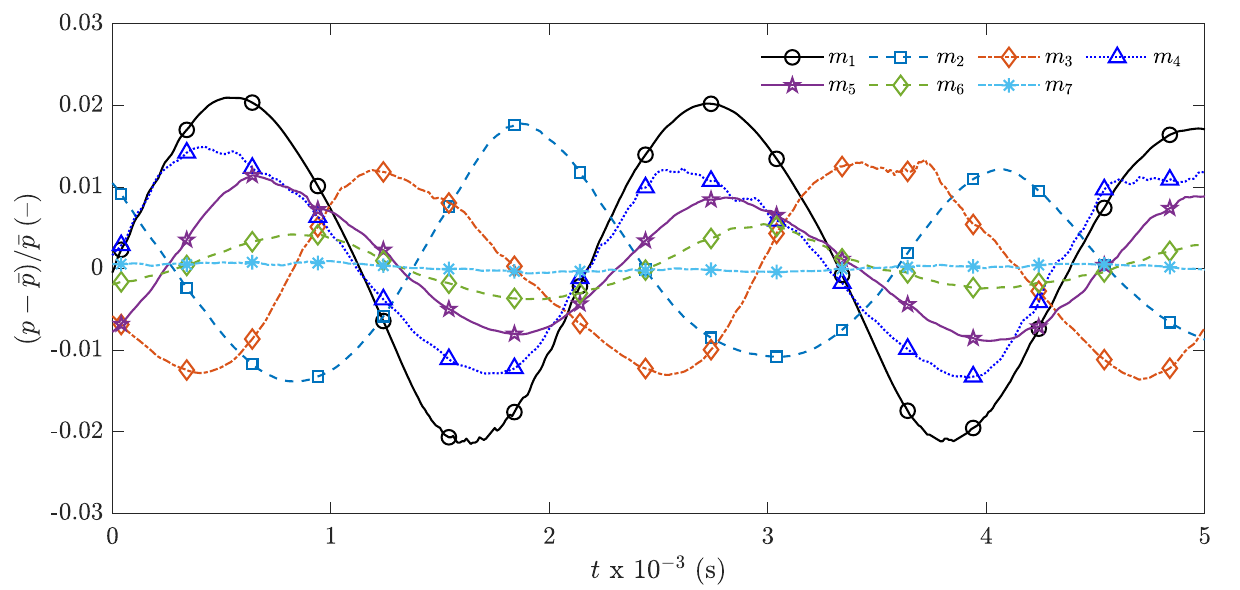}
\caption{LES unsteady pressure pulsations in the wake region.
\label{fig:pressureAmplitude}}
\end{figure*}

 \begin{figure}[ht]
        \centering
        \includegraphics[width=0.8\columnwidth]{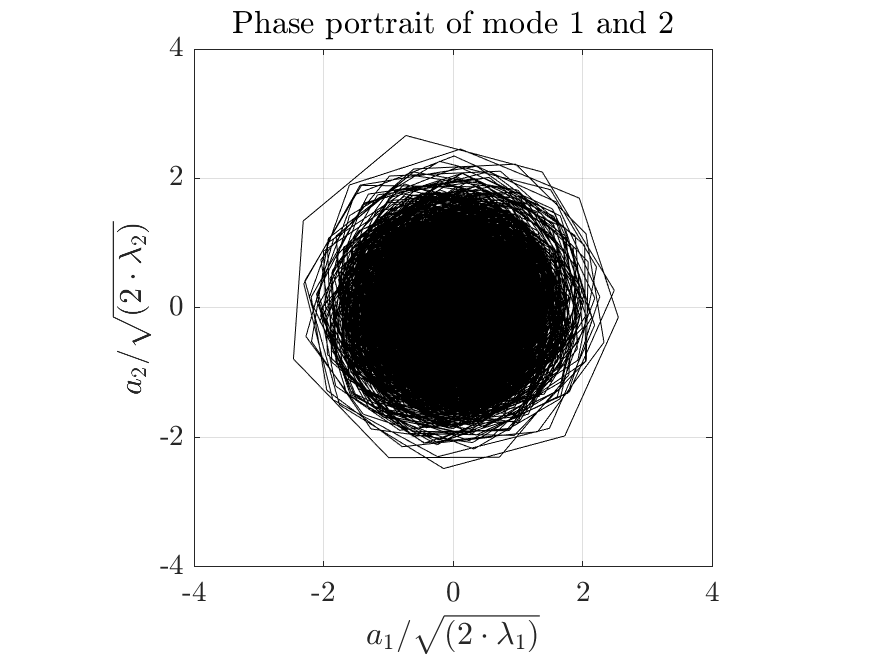} 
        \caption{Phase portrait of temporal coefficient of transverse modes 1 and 2.}
        \label{phase12y}
\end{figure}

\begin{figure}[ht]
\centering
\includegraphics[width=0.5\columnwidth]{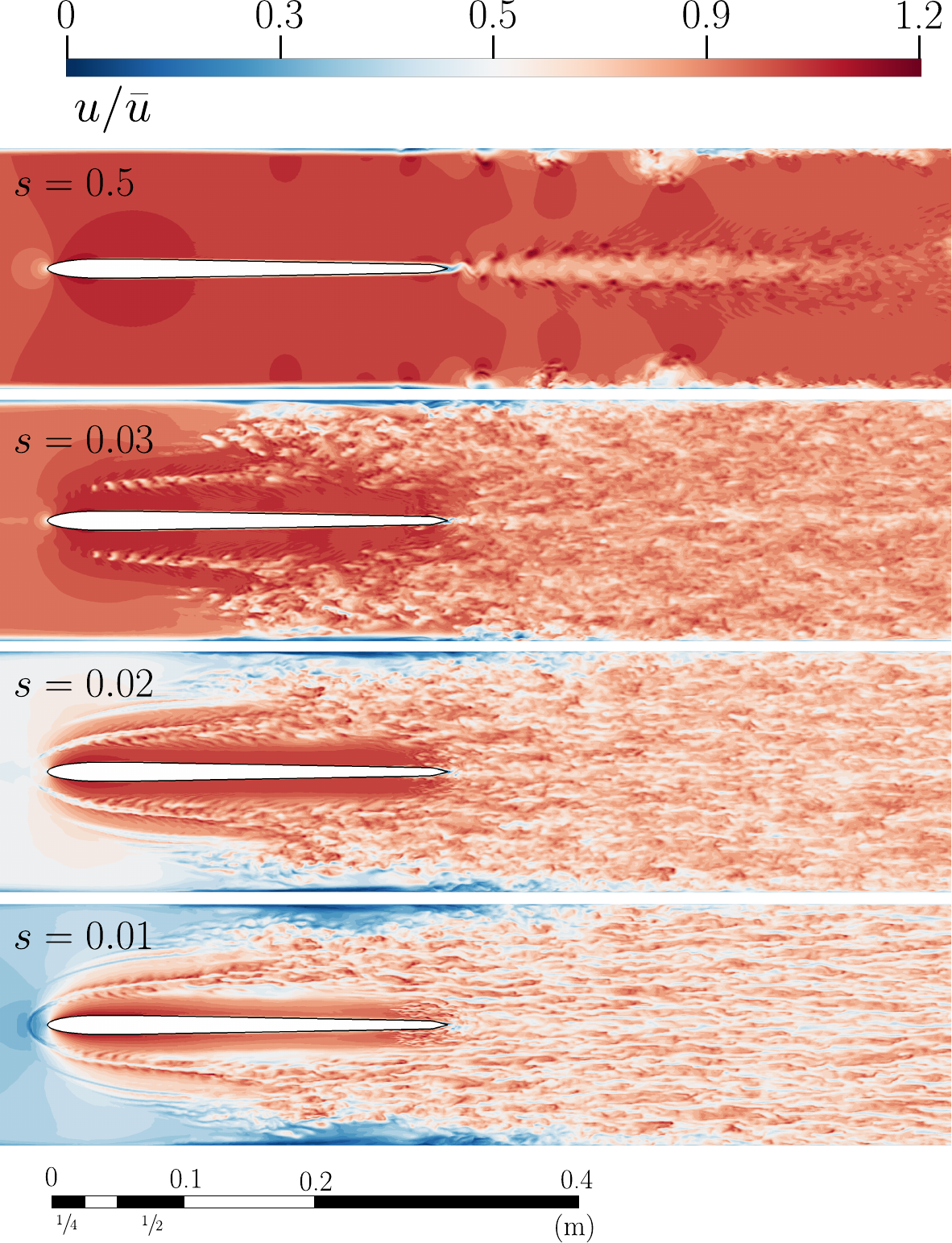}
\caption{Velocity contours at different span values of hydrofoil; $s=0$ and $s=0.5$ indicate the wall and mid-span of the hydrofoil, respectively. \label{fig:velocityPlane2}}
\end{figure}

Figure \ref{S} presents a snapshot of flow velocity measurements from PIV, highlighting the wake region downstream of the hydrofoil's trailing edge. The hydrofoil itself is located out of frame to the left, at the image's mid-span, with water flowing from left to right. The corresponding instantaneous velocity field, derived from the \textit{S} matrix of the previous figure, is shown in Figure \ref{U}. From the complete set of \textit{U} matrices extracted from the PIV recording, all modal components of the flow can be extracted. Figure \ref{mode1} illustrates the most energetic mode, revealing a pattern aligned with the von Kármán vortices, where opposite velocity fields represent the counter-rotating wake vortices. In Figures \ref{mode2} and \ref{mode3}, modes 2 and 3 are identified as \textit{coupled pairs}, characterized by their coherent convective motion and the formation of vortices along the von Kármán vortex streets. These modes exhibit similar energy contributions to the total TKE and display a spatial phase shift of a quarter-wavelength between structures of the same sign, as indicated by the dashed lines in the figures. This observation aligns with the definition of "\textit{coupled pairs}" provided by Riches \cite{riches_proper_2018}. The coupling is further supported by the Power Spectrum Density Function (PSDF), which reveals identical resonance frequency peaks for both modes, as in Figure \ref{psdf2} and \ref{psdf3}. Additional support of the coupling is provided by their phase portraits: Figure \ref{phase12} reveals that modes 1 and 2 are not coupled, as their phase portrait appears chaotic and lacks the smooth circular trajectory indicative of coupling, as it would result in case of an oscillatory behaviour. Conversely, Figure \ref{phase23} shows that modes 2 and 3 are indeed coupled, with a distinct circular phase portrait reflecting oscillatory behavior. Moreover, another evidence of the phase shift has been seen from the quarter-wavelength phase shift seen in the temporal coefficients of the modes (Appendix \ref{APP:POD}, Figure \ref{time23}).

Further analysis of the velocity components in the horizontal ($x$) and transversal ($y$) directions reveals that the POD analysis of the \(U\textsuperscript{x}\) component produces modes nearly identical to the overall velocity modes. In contrast, the transverse component \(U\textsuperscript{y}\) provides valuable insights into the coupling found between the second and third modes. Notably, mode 1, presented in Figure \ref{mode1}, is absent in the transverse component as it is governed by the main horizontal flow direction. The modes 1 and 2 of the transverse components exhibit coupling, as illustrated in Figures \ref{mode1y} and \ref{mode2y}. This coupling is evidenced by multiple factors, including the identical peak frequencies observed in the PSDF analysis, the quarter-wavelength spatial phase shifts (Appendix \ref{APP:POD}, Figure \ref{time12y}), and the circular oscillatory patterns in their phase relationships, as shown in Figure \ref{phase12y}. These findings highlight the coherent dynamics and energy distribution within the transverse modes, further supporting the coupled behavior characteristic of these flow structures. The flow-parallel velocity component \(U\textsuperscript{x} \) is not shown, as it closely matches the modal structure already represented in Figures \ref{mode1}-\ref{mode3}. This consistency is expected, as the primary flow direction dictates the energy distribution, resulting in almost identical modal behaviors between $\tilde{U}^{x}$ and $\tilde{U}$. 

\subsection{Time-dependent vortex shedding}
Large eddy simulations have been conducted at critical Reynolds number with fine mesh of 500 million nodes. To investigate the time dependent vortex shedding and effect of wake, seven monitoring points were created following the trailing edge of hydrofoil. The specific locations of the monitoring points correspond to the profile lines $p_1$, $p_2$, $p_3$, $p_4$, $p_5$, $p_6$, $p_7$ shown in Figure \ref{fig:velprofileloc}, and $y=0$ m. The extracted time dependent pressure variation at the trailing edge is presented in Figure \ref{fig:pressureAmplitude}.

\begin{figure*}[ht]
\centering
\includegraphics[width=1\textwidth]{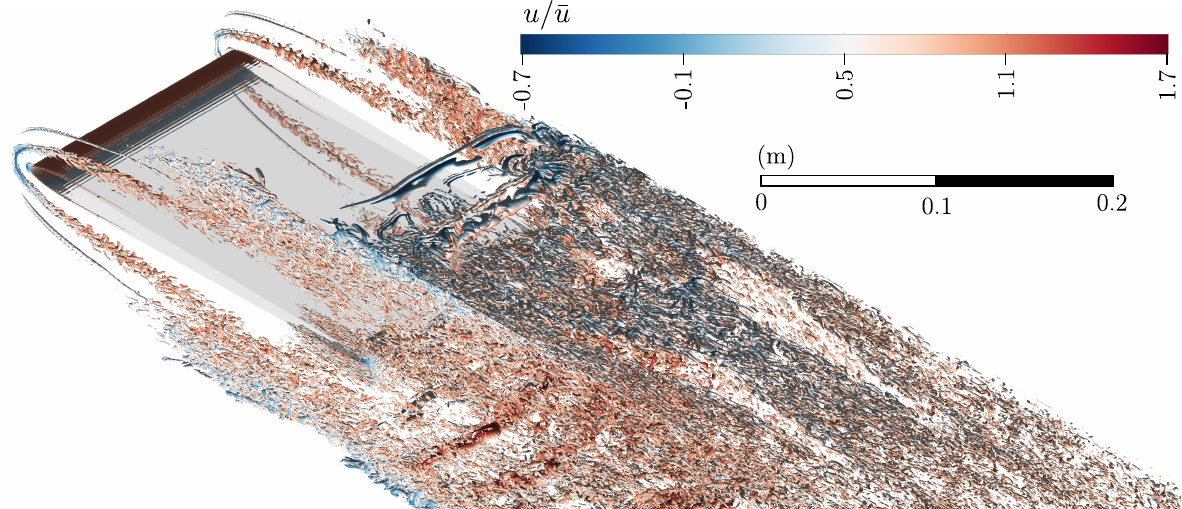}
\caption{Iso-surface of vortex in the hydrofoil test-section, swirling strength = 541 (s{$^{-1}$}). \label{fig:vortexCore}}
\end{figure*}

The monitor point $m_1$ is at the trailing edge and clearly capture the effect of unsteady wake. The time period of pressure amplitude is around 0.0032 second that is equivalent to the shedding frequency of 312 Hz. This is near to the actual measured frequency of 318 Hz from the POD analysis of the transversal velocity component. The peak-to-peak amplitude near the trailing edge is around 2.2\% of the average pressure value. The wake effect is negligible at far downstream of the trailing edge, point $m_7$. Velocity flow field around the hydrofoil is presented in Figure \ref{fig:velocityPlane2}. The contours of instantaneous velocity at four different span ($s$) values, from near wall to the mid span, are studied. At mid span ($s=0.5$), the vortex shedding from the trailing edge is clearly seen. Interestingly, the boundary layer interaction from the upper and lower wall of the test-section is also clearly visible. The boundary layer remains stable and wall attached around the hydrofoil. However, from the trailing edge, the boundary layer appears unstable and breaks apart from the upper and lower wall. At further downstream of the hydrofoil, the vortex shedding diffuses (Figure \ref{vortbreak}), expands, and mixes with the detached eddies of the upper and lower wall boundary layers. The flow field near to the wall (span wise direction) is presented in another velocity contour ($s=0.1$). The flow separation from hydrofoil lead edge corner attached to the span wise wall results in formation and propagation of the vortex. A strong corner vortex expanding downstream from the hydrofoil leading edge is clearly visible. The corner vortex eddies interacts with the boundary layer flow. A three-dimensional iso-surface of the leading edge corner vortex is shown in Figure \ref{fig:vortexCore}. The contours follow the normalized velocity with the velocity at critical Reynolds number. The vortex eddies are filtered with swirling strength of 541 per second. The interaction of the bulk flow, boundary layer separation bubble and trailing edge vortex can be seen clearly. The present numerical model is able to resolve the unsteady eddies very well. Furthermore, the rolling of a large span wise vortex near to the upper wall convecting downstream is also seen. The other span wise vortex at the hydrofoil trailing edge is also resolved and captured by the numerical model.


\clearpage

\section{Conclusions}

This study presented a combined experimental and numerical investigation of the turbulent wake developing downstream of a slender, symmetric hydrofoil with a blunt trailing edge, operating at zero angle of attack and under cavitation-free conditions. High-resolution PIV measurements were compared with scale-resolving numerical simulations based on LES, with the objective of assessing wake dynamics, vortex shedding mechanisms, and the capability of advanced turbulence modeling to reproduce experimentally observed flow features.

The PIV measurements provided detailed, time-resolved velocity fields in the near wake region, revealing a pronounced velocity deficit close to the trailing edge and a gradual recovery toward the bulk flow further downstream. Turbulence intensity profiles clearly identified the development of von Kármán vortex structures along the two distinct, symmetrical paths, with reduced fluctuations at the center due to vortex interaction. Spectral analysis of the transverse velocity component confirmed a dominant vortex shedding frequency on the order of 300 Hz, consistent with prior studies on slender hydrofoils and with the dominant oscillatory modes identified through POD at 318 Hz.

The POD analysis demonstrated that the turbulent kinetic energy is distributed over a large number of modes, reflecting the highly turbulent and broadband nature of the wake. The leading mode was shown to capture the primary wake dynamics, while the second and third modes formed a coupled oscillatory pair associated with von Kármán vortex shedding, as evidenced by their spatial phase shift, identical dominant frequencies, and circular phase portraits. These results confirm the effectiveness of POD in isolating coherent structures embedded in a predominantly stochastic turbulent flow and provide a clear physical interpretation of the wake dynamics measured by PIV.

By combining scale-resolving numerical simulations, including LES, on a very fine mesh of around 500 million nodes successfully resolved both near-wall and wake flow structures. The dominant frequency peak is observed at approximately 314 Hz, which is in close agreement with the frequency identified from the PIV results, considering differences in sampling location and measurement methodology. The comparison between PIV and LES shows that the LES slightly overestimates the momentum deficit relative to PIV, as expected due to domain coverage and resolution differences. The velocity RMS agreement and similar downstream trends validate both the experimental measurements and the LES. These results support the conclusion that LES can reliably reproduce the wake structure and spectral features observed in the experiments, while careful attention must be paid to domain limits when comparing integrated metrics like IMD. In the present work, LES has successfully resolved the leading edge corner vortex and its interaction with the wall separated eddies at downstream.

Overall, the results have shown close agreement between PIV measurements, POD-based modal analysis, and scale-resolving numerical simulations with high-resolution LES. The results provide a validated reference dataset for the assessment of scale-resolving turbulence models applied to hydrofoil wakes and offer new insights into the wake physics of slender symmetric hydrofoils under high-Reynolds-number conditions. These findings are relevant to the prediction and mitigation of vortex-induced vibrations in hydraulic machinery and contribute to the development of more reliable design and analysis tools for hydrofoil-based components

\section*{Acknowledgment}
The computational resources used under the e-infrastructure project No. NN9504K --- Developing the next-generation of hydropower technology. The resources were provided by Sigma2 --- the National Infrastructure for High-Performance Computing and Data Storage in Norway.

\section*{Nomenclature}
$A$ = POD time-coefficient matrix\\
$c$ = hydrofoil chord length (m)\\
$CFD$ = Computational Fluid Dynamics\\
$D_h$ = hydraulic diameter of test section (m)\\
$d$ = characteristic length (m)\\
$f$ = frequency (Hz)\\
$IMD$ = Integrated Momentum Deficit (m)\\
$LES$ = Large Edyy Simulation\\
$p_i$ = profile location downstream of trailing edge (m)\\
$P$ = pressure (Pa)\\
$PIV$ = Particle Image Velocimetry\\
$S$ = velocity matrix\\
$Q$ = volumetric flow rate (m$^3$ s$^{-1}$)\\
$t$ = time (s)\\
$TKE$ = turbulent kinetic energy (m$^2$ s$^{-2}$)\\
$U$ = fluctuating velocity matrix\\
$U_\infty$ = bulk velocity (m s$^{-1}$)\\
$u'$ = streamwise velocity fluctuation (m s$^{-1}$)

\textbf{Greek symbols}\\
$\lambda_k$ = eigenvalue of POD mode $k$ \\
$\nu$ = kinematic viscosity (m$^2$ s$^{-1}$)\\
$\rho$ = fluid density (kg m$^{-3}$)\\
$\Phi$ = matrix of POD spatial modes \\

\textbf{Dimensionless quantities and groups}\\
$a_k$ = temporal coefficient of POD mode $k$ (-)\\
$m$ = number of PIV snapshots (-)\\
$n$ = number of spatial points per snapshot (-)\\
$Re$ = Reynolds number (-)\\
$St$ = Strouhal number, $St = f d / U_\infty$

\textbf{Subscripts and superscripts}\\
$b$ = bulk value\\
$\infty$ = free-stream value\\
$'$ = fluctuating quantity\\
$k$ = POD mode index\\
$x$ = streamwise direction\\
$y$ = transversal direction

\clearpage

\begin{appendices}
\section{LES Velocity profiles}       
\label{APP:LES}

\begin{figure}[ht]
\centering
    \includegraphics[width=0.5\columnwidth]{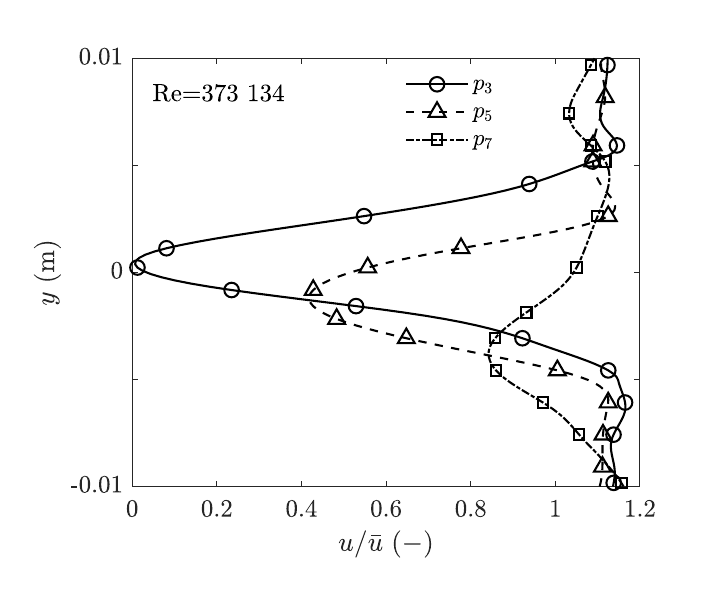}
    \caption{\(Re = 3.73 \times 10^{5}\).}
    \label{fig:velprofile3ms}
\end{figure}

\begin{figure}[ht]
\centering
    \includegraphics[width=\columnwidth]{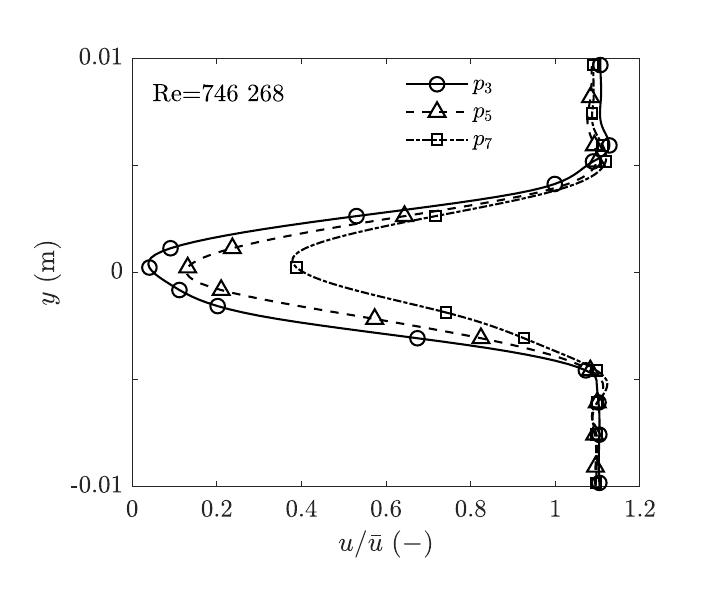}
    \caption{\(Re = 7.46 \times 10^{5}\).}
    \label{fig:velprofile5ms}
\end{figure}

\begin{figure}[ht]
\centering
    \includegraphics[width=0.9\columnwidth]{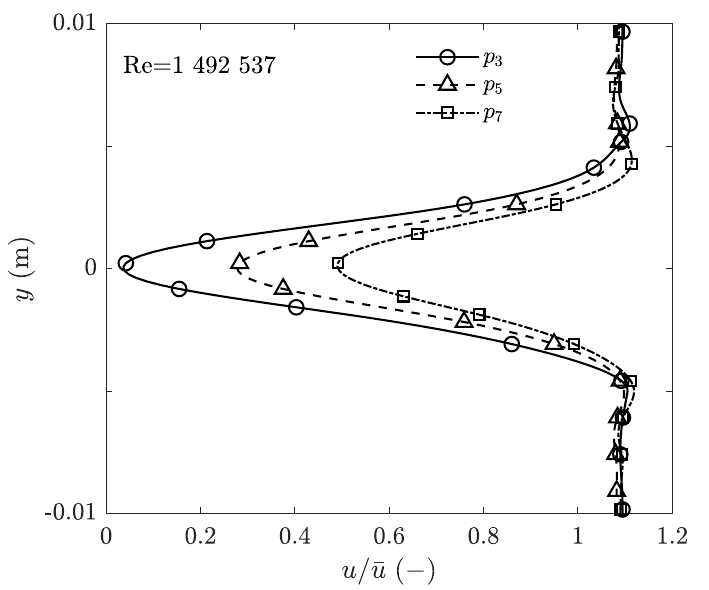}
    \caption{\(Re = 1.49 \times 10^{6}\).}
    \label{fig:velprofile10ms}
    \end{figure}

\begin{figure}[ht]
\centering
    \includegraphics[width=0.9\columnwidth]{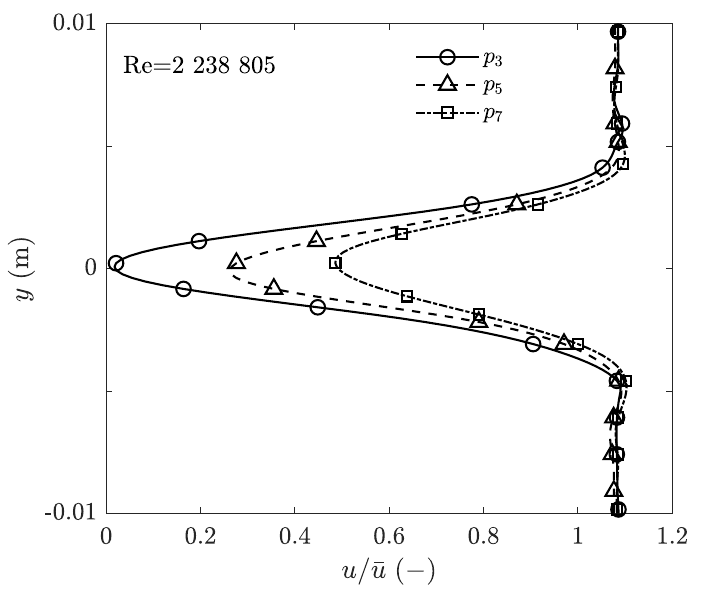}
    \caption{\(Re = 2.24 \times 10^{6}\).}
    \label{fig:velprofile15ms}
\end{figure}

\section{POD additional results}
\label{APP:POD}
\begin{figure}[ht]
        \centering
        \includegraphics[width=\columnwidth]{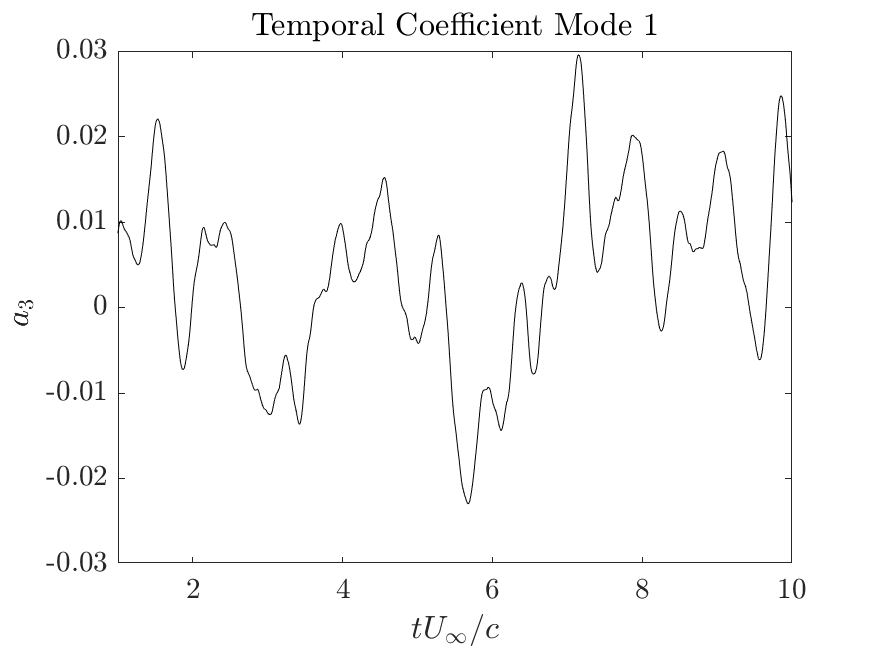} 
        \caption{Temporal coefficient of mode 1.}
        \label{time1}
\end{figure}

\begin{figure}[ht]
        \centering
        \includegraphics[width=\columnwidth]{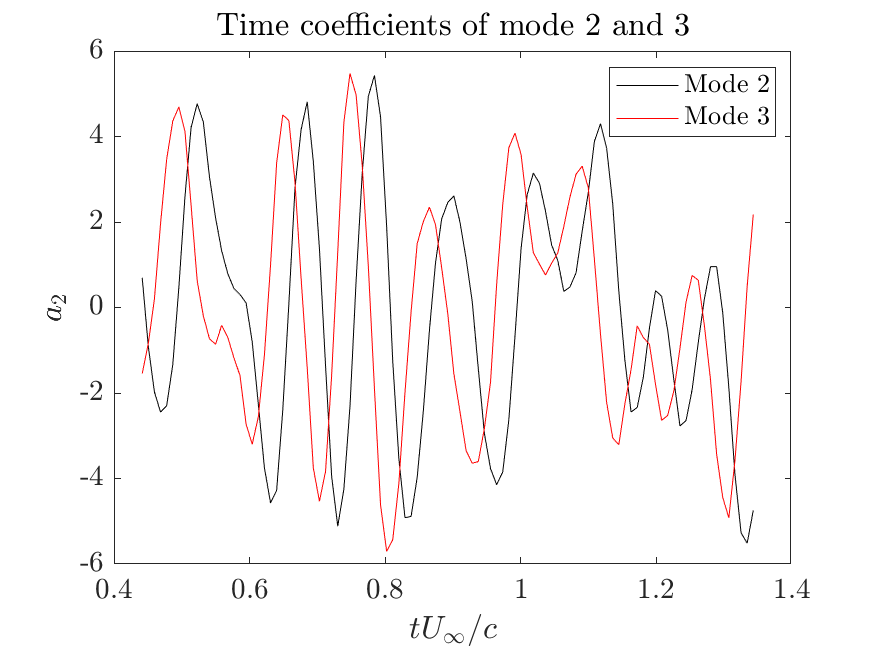} 
        \caption{Detail of the temporal coefficient of mode 2 and 3.}
        \label{time23}
\end{figure}

\begin{figure}[ht]
    \centering
    \includegraphics[width=\columnwidth]{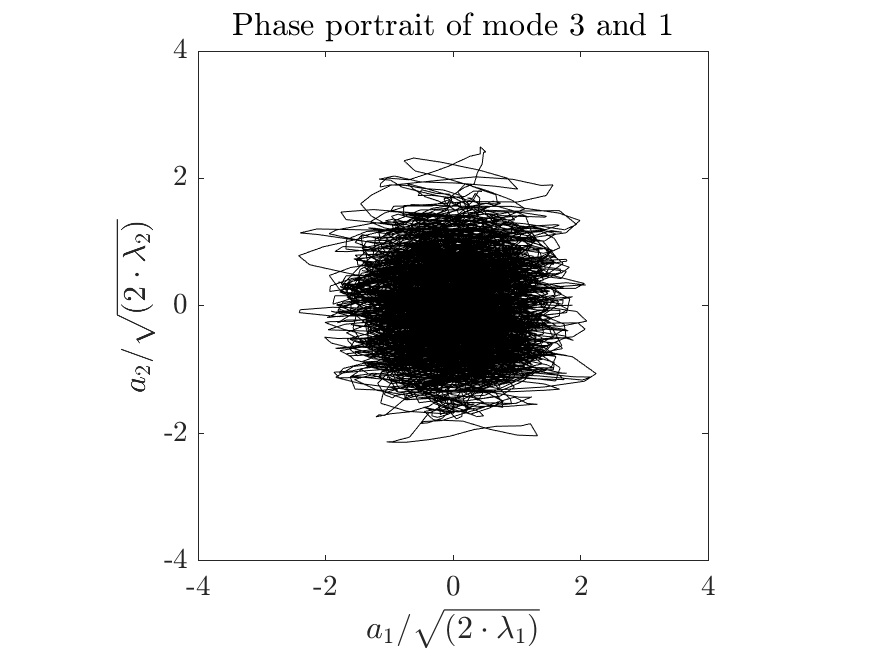} 
    \caption{Phase portrait of temporal coefficient of modes 1 and 3.}
    \label{phase23y}
\end{figure}

\begin{figure}[ht]
    \centering
    \includegraphics[width=\columnwidth]{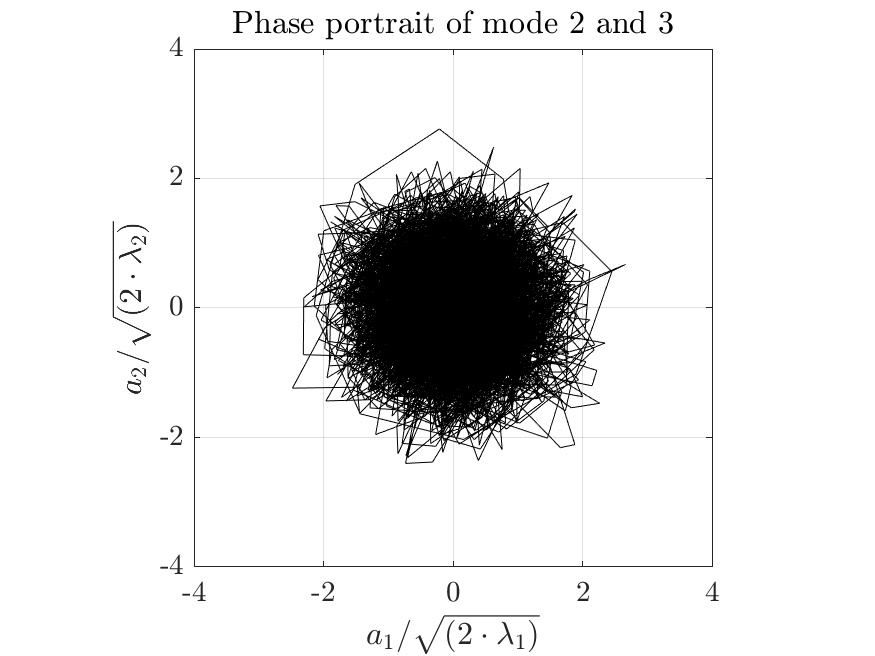} 
    \caption{Phase portrait of temporal coefficient of transverse modes 2 and 3.}
    \label{phase23y}
\end{figure}

\begin{figure}[ht]
    \centering
    \includegraphics[width=\columnwidth]{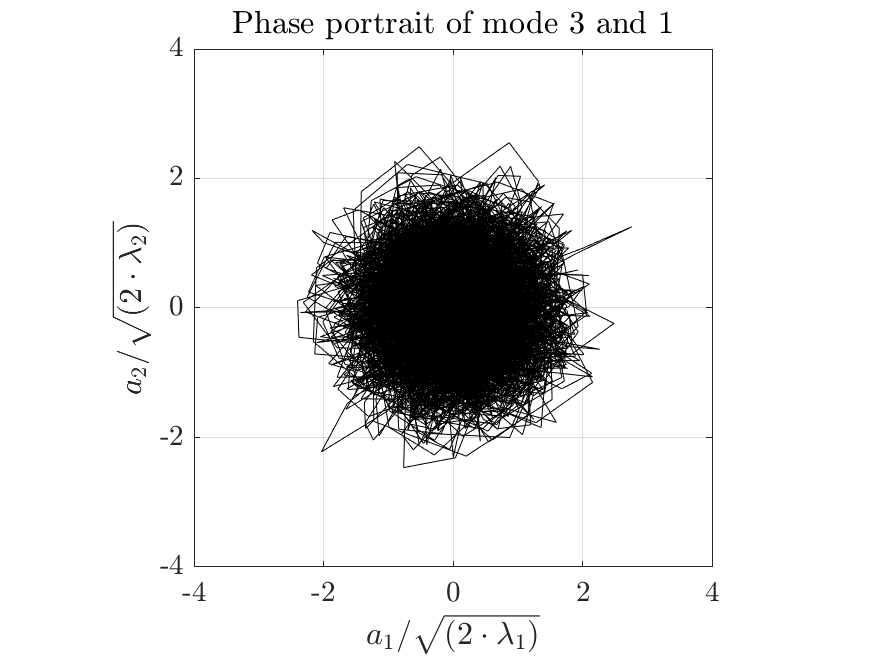} 
    \caption{Phase portrait of temporal coefficients of transverse modes 1 and 3.}
    \label{phase31y}
\end{figure}

\begin{figure}[ht]
    \centering
    \includegraphics[width=\columnwidth]{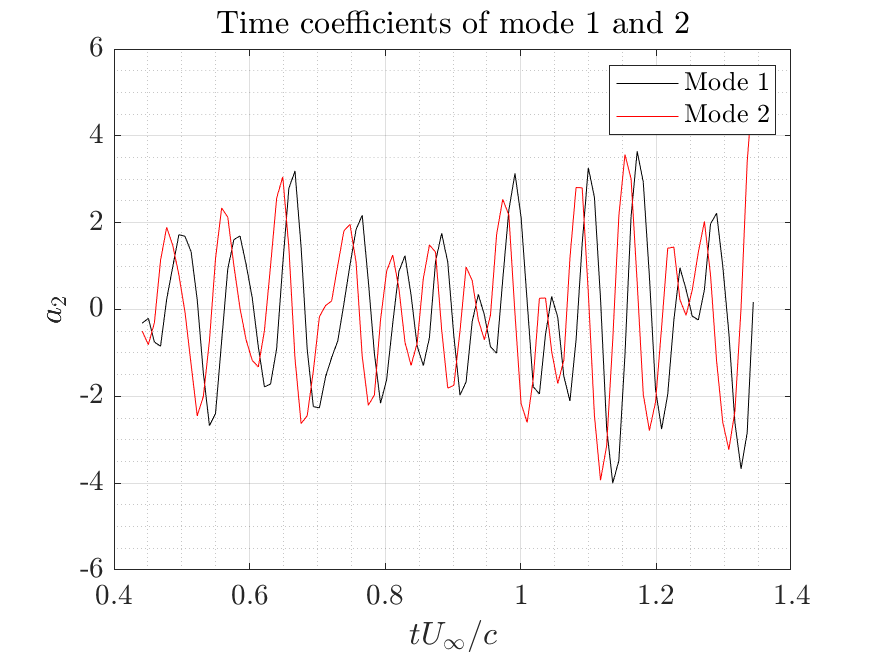} 
    \caption{Detail of the temporal coefficient of transverse mode 1 and 2.}
    \label{time12y}
\end{figure}

\end{appendices}

\clearpage
\nocite{*}
\bibliographystyle{asmejour}
\bibliography{Bibliography}

@phdthesis{zobeiri_effect_2012,
	title = {Effect of hydrofoil trailing edge geometry on the wake dynamics},
	copyright = {openaccess},
	url = {https://infoscience.epfl.ch/handle/20.500.14299/71148},
	language = {en},
	urldate = {2026-01-29},
	school = {Lausanne, EPFL},
	author = {Zobeiri, Amirreza},
	year = {2012},
	doi = {10.5075/EPFL-THESIS-5218},
}

@mastersthesis{arunn_master,
    author 	    = {Arunn Kamalaraja},
    title 	    = {Investigation of Corner Vortex in Radial Cascade and Interaction with the Trailing Edge Vortex},
    school 	    = {Norwegian University of Science and Technology (NTNU)},
    address 	= {Trondheim, Norway},
    year 		= {2021},
    url         = {https://hdl.handle.net/11250/2779968}
}

@article{ben-gida_stratified_2016,
	title = {A stratified wake of a hydrofoil accelerating from rest},
	volume = {70},
	issn = {08941777},
	doi = {10.1016/j.expthermflusci.2015.09.031},
	journal = {Experimental Thermal and Fluid Science},
	author = {Ben-Gida, Hadar and Liberzon, Alex and Gurka, Roi},
	month = jan,
	year = {2016},
	pages = {366--380},
}

@article{chen_numerical_2020,
	title = {Numerical investigation of the cavitation effects on the vortex shedding from a hydrofoil with blunt trailing edge},
	volume = {5},
	issn = {2311-5521},
	doi = {10.3390/fluids5040218},
	number = {4},
	urldate = {2026-01-29},
	journal = {Fluids},
	author = {Chen, Jian and Geng, Linlin and Escaler, Xavier},
	month = nov,
	year = {2020},
	pages = {218},
}

@article{dauengauer_flow_2018,
	title = {Flow around a low-aspect-ratio wall-bounded {2D} hydrofoil: a {LES}/{PIV} study},
	volume = {1128},
	issn = {1742-6588, 1742-6596},
	doi = {10.1088/1742-6596/1128/1/012085},
	urldate = {2026-01-29},
	journal = {Journal of Physics: Conference Series},
	author = {Dauengauer, E I and Mullyadzhanov, R I and Timoshevskiy, M V and Zapryagaev, I I and Pervunin, K S},
	month = nov,
	year = {2018},
	pages = {012085},
}

@article{hu_numerical_2020,
	title = {Numerical simulation on vortex shedding from a hydrofoil in steady flow},
	volume = {8},
	issn = {2077-1312},
	doi = {10.3390/jmse8030195},
	number = {3},
	urldate = {2026-01-29},
	journal = {Journal of Marine Science and Engineering},
	author = {Hu, Jian and Wang, Zibin and Zhao, Wang and Sun, Shili and Sun, Cong and Guo, Chunyu},
	month = mar,
	year = {2020},
	pages = {195},
}

@article{afgan_large_2011,
	title = {Large eddy simulation of the flow around single and two side-by-side cylinders at subcritical {Reynolds} numbers},
	volume = {23},
	issn = {1070-6631, 1089-7666},
	url = {http://aip.scitation.org/doi/10.1063/1.3596267},
	doi = {10.1063/1.3596267},
	language = {en},
	number = {7},
	urldate = {2022-06-10},
	journal = {Physics of Fluids},
	author = {Afgan, I. and Kahil, Y. and Benhamadouche, S. and Sagaut, P.},
	month = jul,
	year = {2011},
	pages = {075101},
}

@article{chen_experimental_2017,
	title = {Experimental study on the multimodal dynamics of the turbulent horseshoe vortex system around a circular cylinder},
	volume = {29},
	issn = {1070-6631, 1089-7666},
	url = {http://aip.scitation.org/doi/10.1063/1.4974523},
	doi = {10.1063/1.4974523},
	language = {en},
	number = {1},
	urldate = {2022-06-10},
	journal = {Physics of Fluids},
	author = {Chen, Qigang and Qi, Meilan and Zhong, Qiang and Li, Danxun},
	month = jan,
	year = {2017},
	pages = {015106},
}

@article{del_pino_structure_2011,
	title = {Structure of trailing vortices: {Comparison} between particle image velocimetry measurements and theoretical models},
	volume = {23},
	issn = {1070-6631, 1089-7666},
	shorttitle = {Structure of trailing vortices},
	url = {http://aip.scitation.org/doi/10.1063/1.3537791},
	doi = {10.1063/1.3537791},
	language = {en},
	number = {1},
	urldate = {2022-06-10},
	journal = {Physics of Fluids},
	author = {del Pino, C. and Parras, L. and Felli, M. and Fernandez-Feria, R.},
	month = jan,
	year = {2011},
	pages = {013602},
}

@article{temmerman_investigation_2003,
	title = {Investigation of wall-function approximations and subgrid-scale models in large eddy simulation of separated flow in a channel with streamwise periodic constrictions},
	volume = {24},
	copyright = {https://www.elsevier.com/tdm/userlicense/1.0/},
	issn = {0142727X},
	url = {https://linkinghub.elsevier.com/retrieve/pii/S0142727X02002229},
	doi = {10.1016/S0142-727X(02)00222-9},
	language = {en},
	number = {2},
	urldate = {2026-01-29},
	journal = {International Journal of Heat and Fluid Flow},
	author = {Temmerman, Lionel and Leschziner, Michael A. and Mellen, Christopher P. and Fröhlich, Jochen},
	month = apr,
	year = {2003},
	pages = {157--180},
}

@article{heskestad_influence_1960,
	title = {Influence of trailing-edge geometry on hydraulic-turbine-blade vibration resulting from vortex excitation},
	volume = {82},
	issn = {0022-0825},
	url = {https://asmedigitalcollection.asme.org/gasturbinespower/article/82/2/103/404982/Influence-of-TrailingEdge-Geometry-on},
	doi = {10.1115/1.3672718},
	number = {2},
	urldate = {2026-01-29},
	journal = {Journal of Engineering for Power},
	author = {Heskestad, Gunnar and Olberts, D. R.},
	month = apr,
	year = {1960},
	pages = {103--109},
}

@article{sagmo_piv_2019,
	title = {{PIV} measurements and {CFD} simulations of a hydrofoil at lock-in},
	volume = {240},
	copyright = {http://iopscience.iop.org/info/page/text-and-data-mining},
	issn = {1755-1315},
	url = {https://iopscience.iop.org/article/10.1088/1755-1315/240/6/062006},
	doi = {10.1088/1755-1315/240/6/062006},
	urldate = {2026-01-29},
	journal = {IOP Conference Series: Earth and Environmental Science},
	author = {Sagmo, K F and Tengs, E O and Bergan, C W and Storli, P T},
	month = mar,
	year = {2019},
	pages = {062006},
}

@article{besirovic_vortex_2020,
	title = {Vortex generator’s effect on trailing edge vortex shedding and fluid structure interaction},
	volume = {1608},
	copyright = {http://iopscience.iop.org/info/page/text-and-data-mining},
	issn = {1742-6588, 1742-6596},
	url = {https://iopscience.iop.org/article/10.1088/1742-6596/1608/1/012002},
	doi = {10.1088/1742-6596/1608/1/012002},
	urldate = {2026-01-29},
	journal = {Journal of Physics: Conference Series},
	author = {Besirovic, H and Sagmo, K F and Storli, P T},
	month = aug,
	year = {2020},
	pages = {012002},
}

@article{williamson_vortex_1996,
	title = {Vortex {Dynamics} in the {Cylinder} {Wake}},
	volume = {28},
	issn = {0066-4189, 1545-4479},
	url = {https://www.annualreviews.org/doi/10.1146/annurev.fl.28.010196.002401},
	doi = {10.1146/annurev.fl.28.010196.002401},
	language = {en},
	number = {1},
	urldate = {2026-01-29},
	journal = {Annual Review of Fluid Mechanics},
	author = {Williamson, C H K},
	month = jan,
	year = {1996},
	pages = {477--539},
}

@article{westerweel_fundamentals_1997,
	title = {Fundamentals of digital particle image velocimetry},
	volume = {8},
	issn = {0957-0233, 1361-6501},
	doi = {10.1088/0957-0233/8/12/002},
	number = {12},
	urldate = {2026-01-29},
	journal = {Measurement Science and Technology},
	author = {Westerweel, J},
	month = dec,
	year = {1997},
	pages = {1379--1392},
}

@book{westerweel_digital_1993,
	address = {Delft},
	title = {Digital particle image velocimetry: {Theory} and application},
	isbn = {9789062758814},
	shorttitle = {Digital particle image velocimetry},
	language = {eng},
	publisher = {Delft University Press},
	author = {Westerweel, Jerry},
	year = {1993},
}

@book{raffel_particle_2018,
	address = {Cham},
	edition = {3rd ed},
	title = {Particle image velocimetry: a practical guide},
	isbn = {9783319688527},
	publisher = {Springer International Publishing AG},
	author = {Raffel, Markus},
	collaborator = {Willert, Christian E. and Scarano, Fulvio and Kähler, Christian J. and Wereley, Steve T. and Kompenhans, Jürgen},
	year = {2018},
}

@article{wieneke_piv_2015,
	title = {{PIV} uncertainty quantification from correlation statistics},
	volume = {26},
	issn = {0957-0233, 1361-6501},
	doi = {10.1088/0957-0233/26/7/074002},
	number = {7},
	urldate = {2026-01-29},
	journal = {Measurement Science and Technology},
	author = {Wieneke, Bernhard},
	month = jul,
	year = {2015},
	pages = {074002},
}

@article{lumley_toward_1970,
	title = {Toward a turbulent constitutive relation},
	volume = {41},
	issn = {0022-1120, 1469-7645},
	doi = {10.1017/S0022112070000678},
	number = {2},
	urldate = {2026-01-29},
	journal = {Journal of Fluid Mechanics},
	author = {Lumley, J. L.},
	month = apr,
	year = {1970},
	pages = {413--434},
}

@article{sirovich_turbulence_1987,
	title = {Turbulence and the dynamics of coherent structures. {I}. {Coherent} structures},
	volume = {45},
	issn = {0033-569X, 1552-4485},
	doi = {10.1090/qam/910462},
	language = {en},
	number = {3},
	urldate = {2026-01-29},
	journal = {Quarterly of Applied Mathematics},
	author = {Sirovich, Lawrence},
	month = oct,
	year = {1987},
	pages = {561--571},
}

@inproceedings{weiss_tutorial_2019,
	address = {Dallas, Texas},
	title = {A tutorial on the proper orthogonal decomposition},
	isbn = {9781624105890},
	doi = {10.2514/6.2019-3333},
	urldate = {2026-01-29},
	booktitle = {{AIAA} {Aviation} 2019 {Forum}},
	publisher = {American Institute of Aeronautics and Astronautics},
	author = {Weiss, Julien},
	month = jun,
	year = {2019},
}

@article{fiedler_coherent_1988,
	title = {Coherent structures in turbulent flows},
	volume = {25},
	issn = {03760421},
	doi = {10.1016/0376-0421(88)90001-2},
	number = {3},
	urldate = {2026-01-29},
	journal = {Progress in Aerospace Sciences},
	author = {Fiedler, H.E.},
	month = jan,
	year = {1988},
	pages = {231--269},
}

@article{menter_review_2009,
	title = {Review of the shear-stress transport turbulence model experience from an industrial perspective},
	volume = {23},
	issn = {1061-8562},
	doi = {10.1080/10618560902773387},
	number = {4},
	journal = {International Journal of Computational Fluid Dynamics},
	author = {Menter, F R},
	year = {2009},
	pages = {305--316},
}

@article{menter_scale-adaptive_2010,
	title = {The scale-adaptive simulation method for unsteady turbulent flow predictions. {Part} 1: {Theory} and model description},
	volume = {85},
	issn = {1386-6184},
	doi = {10.1007/s10494-010-9264-5},
	number = {1},
	journal = {Flow, Turbulence and Combustion},
	author = {Menter, F R and Egorov, Y},
	year = {2010},
	pages = {113--138},
}

@article{egorov_scale-adaptive_2010,
	title = {The scale-adaptive simulation method for unsteady turbulent flow predictions. {Part} 2: {Application} to complex flows},
	volume = {85},
	issn = {1386-6184},
	doi = {10.1007/s10494-010-9265-4},
	number = {1},
	journal = {Flow, Turbulence and Combustion},
	author = {Egorov, Y and Menter, F R and Lechner, R and Cokljat, D},
	year = {2010},
	pages = {139--165},
}

@article{riches_proper_2018,
	title = {Proper orthogonal decomposition analysis of a circular cylinder undergoing vortex-induced vibrations},
	volume = {30},
	issn = {1070-6631, 1089-7666},
	doi = {10.1063/1.5046090},
	number = {10},
	urldate = {2026-01-29},
	journal = {Physics of Fluids},
	author = {Riches, Graham and Martinuzzi, Robert and Morton, Chris},
	month = oct,
	year = {2018},
	pages = {105103},
}

@article{fernandez-aldama_characterization_2025,
	title = {Characterization of vortex-shedding regimes and lock-in response of a wind turbine airfoil with two high-fidelity simulation approaches},
	volume = {10},
	issn = {2366-7451},
	doi = {10.5194/wes-10-17-2025},
	number = {1},
	urldate = {2026-01-29},
	journal = {Wind Energy Science},
	author = {Fernandez-Aldama, Ricardo and Papadakis, George and Lopez-Garcia, Oscar and Avila-Sanchez, Sergio and Riziotis, Vasilis A. and Cuerva-Tejero, Alvaro and Gallego-Castillo, Cristobal},
	month = jan,
	year = {2025},
	pages = {17--39},
}

@article{ausoni_cavitation_2007,
	title = {Cavitation influence on von kármán vortex shedding and induced hydrofoil vibrations},
	volume = {129},
	issn = {0098-2202, 1528-901X},
	doi = {10.1115/1.2746907},
	number = {8},
	urldate = {2026-01-29},
	journal = {Journal of Fluids Engineering},
	author = {Ausoni, Philippe and Farhat, Mohamed and Escaler, Xavier and Egusquiza, Eduard and Avellan, François},
	month = aug,
	year = {2007},
	pages = {966--973},
}

@article{siala_characterization_2016,
	title = {Characterization of vortex dynamics in the near wake of an oscillating flexible foil},
	volume = {138},
	issn = {0098-2202, 1528-901X},
	doi = {10.1115/1.4033959},
	number = {10},
	urldate = {2026-01-29},
	journal = {Journal of Fluids Engineering},
	author = {Siala, Firas F. and Totpal, Alexander D. and Liburdy, James A.},
	month = oct,
	year = {2016},
	pages = {101202},
}

@inproceedings{ziazi_tomographic_2019,
	address = {San Francisco, California, USA},
	title = {A {Tomographic} {PIV} {Study} and {Comparison} of {Vortex} {Identification} {Methods} on {NACA} 63-215 {Hydrofoil} {Wake} {Structure}},
	isbn = {9780791859070},
	doi = {10.1115/AJKFluids2019-5550},
	urldate = {2026-01-29},
	booktitle = {Volume 4: {Fluid} {Measurement} and {Instrumentation}; {Micro} and {Nano} {Fluid} {Dynamics}},
	publisher = {American Society of Mechanical Engineers},
	author = {Ziazi, Reza M. and Goudarzi, Navid},
	month = jul,
	year = {2019},
	pages = {V004T04A014},
}

@article{fruman_effect_1995,
	title = {Effect of hydrofoil planform on tip vortex roll-up and cavitation},
	volume = {117},
	issn = {0098-2202, 1528-901X},
	doi = {10.1115/1.2816806},
	number = {1},
	urldate = {2026-01-29},
	journal = {Journal of Fluids Engineering},
	author = {Fruman, D. H. and Cerrutti, P. and Pichon, T. and Dupont, P.},
	month = mar,
	year = {1995},
	pages = {162--169},
}

\end{document}